\pdfoutput=1 
\documentclass[11pt]{article}
\usepackage{amsmath}
\usepackage{amssymb}
\usepackage{slashed}
\usepackage{mathtools}
\usepackage{graphicx,epstopdf}
\usepackage{cite}
\usepackage{pbox}
\usepackage{geometry}
\usepackage{float}
\usepackage{array}
\usepackage{amsfonts}
\usepackage{braket}
\usepackage{graphicx}
\usepackage[font={small}]{caption}
\usepackage[table]{xcolor}
\usepackage{xcolor}
\usepackage{hyperref}
\hypersetup{
    colorlinks=true,
    linkcolor=red,
    filecolor=magenta,      
    urlcolor=blue,
   citecolor=blue,
} 

\providecommand{\xlink}[1]
  {\href{http://arxiv.org/abs/#1}{arXiv:#1}}
\definecolor{RawSienna}{cmyk}{0,0.72,1,0.45}
\definecolor{dgreen}{rgb}{0.0,0.42,0.13}
\definecolor{darkblue}{rgb}{0.0, 0.0, 0.55}
\definecolor{cornellred}{rgb}{0.7, 0.11, 0.11}
\definecolor{calpolypomonagreen}{rgb}{0.08, 0.5, 0.5}

\def\beq{\begin{equation}}
\def\eeq{\end{equation}}
\def\bea{\begin{eqnarray}}
\def\eea{\end{eqnarray}}
\makeatletter
\@addtoreset{equation}{section}

\begin{document}
\title{\LARGE \bf Importance of generalized $\mu\tau$ symmetry and its CP extension on neutrino mixing and leptogenesis }
\author{{\bf Rome Samanta$^1$\footnote{R.Samanta@soton.ac.uk}, Roopam Sinha$^2$\footnote{roopam.sinha@saha.ac.in}, Ambar Ghosal$^2$\footnote{ambar.ghosal@saha.ac.in},}\\
1. Physics and Astronomy,
University of Southampton,
Southampton, SO17 1BJ, U.K.\\2. Saha Institute of Nuclear Physics, HBNI, 1/AF Bidhannagar, Kolkata 700064, India}
\maketitle
\begin{abstract} 
Within the framework of residual symmetry, two $\mathbb{Z}_2$ type associate $\mu\tau$ interchange symmetries robustly constrain the Dirac CP phase $\delta$ in a model independent way. Both of them predict simultaneous maximality of  $\delta$ and  the atmospheric  mixing angle  $\theta_{23}$.  We show how these well known correlations will be changed if we generalize the $\mu\tau$ interchange symmetry to a $\mu\tau$ mixing symmetry. In particular, we show that the stringent condition of simultaneous maximality could be relaxed even with a very small departure from the exact $\mu\tau$ interchange. In addition, the present neutrino data on $\delta$ and $\theta_{23}$ can be  explained better by  the mixing symmetry.  After discussing the impact of the $\mu\tau$ mixing in some realistic neutrino mass models, we show how the proposed mixing could be realized with two simultaneous CP transformations which also lead to  novel and testable correlations between $\delta$ and the mixing angles $\theta_{ij}$. Next we discuss in particular, the `three flavour regime' of leptogenesis within the CP extended framework and show, unlike the ordinary CP extended $\mu\tau$ interchange symmetry, a resonant leptogenesis is possible due the generalization of $\mu\tau$ interchange to the $\mu\tau$ mixing and the resulting baryon asymmetry always requires a nonmaximal $\theta_{23}$ owing to the fact that the baryon to photon ratio $\eta_B$ vanishes in the exact limit of $\theta_{23}=\pi/4$. This is one of the   robust predictions of this framework. The CP extended $\mu\tau$ mixing is also a novel example of a low energy effective model that provides an important insight to the off-diagonal terms of the flavour coupling matrix which have usually been neglected in literature to compute the final baryon asymmetry, in particular in the models with flavour symmetries. 
\end{abstract} 

\section{Introduction} In the present era of precision measurement of neutrino oscillation parameters such as the three mixing angles and two independent mass-squared differences,  yet unsettled issues like the mass ordering, octant of the atmospheric mixing angle $\theta_{23}$ and the value of the Dirac CP phase $\delta$ have also drawn a lot of attention. Precise determination of the first two, for example, would confront many of the Grand Unified Theories (GUT)\cite{G1,G2,G3,G4,G5,G6,G7,G8,G9,G10} of neutrino masses and mixing while a CP violating value of  $\delta$ would have an immense implication on the observed dominance of  matter over the antimatter\cite{L1,L2,L2p,L2pp,L3,L4,Dolan:2018qpy}. Latest combined global analysis of experiments such as T2K\cite{Abe:2017bay}, NO$\nu$A\cite{Adamson:2017qqn}, MINOS\cite{minos} and RENO\cite{reno} favour a Normal Mass Ordering (NMO) at 2$\sigma$\cite{nufit,old} and shows a preference for the second octant of  $\theta_{23}$ with a best-fit $\sin^2\theta_{23}=0.58$. On the other hand, best fits for the phase $\delta$ are close to $284^\circ$ for an Inverted Mass Ordering (IMO) and $215^\circ$ for a NMO while the CP conserving values (i.e., $\delta=0,\pi$) as well as one of the CP violating value $\delta=\pi/2$ are disfavoured at 70$\%$ CL and 99$\%$ CL respectively\cite{nufit}. For the NMO with  solar  and atmospheric mass-squared differences as $\Delta m_{12}^2=7.39^{+0.21}_{-0.20}\times10^{-5}{\rm eV}^2$  and $\Delta m_{31}^2=2.52^{+0.033}_{-0.032}\times10^{-3}{\rm eV}^2$, global fit values of the three mixing angles and the CP phase $\delta$ are summarized  in TABLE \ref{t1}.
\begin{table}[H]
\caption{Best-fit, 1$\sigma$ and 3$\sigma$ ranges of three mixing angles and the Dirac CP phase $\delta$ for NMO (\href{http://www.nu-fit.org/?q=node/12}{NuFIT\cite{nufit}})}\label{t1}
\hspace{2cm}
\begin{tabular}{|c|c|}
\hline
&$\theta_{12}/^\circ$~~~~~~~~~~~$\theta_{23}/^\circ$~~~~~~~~~~~$\theta_{13}/^\circ$~~~~~~~~~~~$\delta/^\circ$\\
\hline
\hline
${\rm bf}\pm1\sigma$&$33.82^{+0.78}_{-0.76}$~~~~~~~$49.6^{+1.0}_{-1.2}$~~~~~~~$8.61^{+0.13}_{-0.13}$~~~~~~~$215^{+40}_{-29}$\\
\hline
$3\sigma$&$31.61\rightarrow 36.27$~~~$40.3\rightarrow 52.4$~~~$8.22\rightarrow 8.99$~~~$125\rightarrow 392$\\
\hline
\end{tabular}
\end{table}

Thus, it is a crucial juncture in regard to the measurement of $\delta$ as well as $\theta_{23}$, since many of the well-acknowledged neutrino mass models that predict, e.g.,  a co-bimaximal mixing ($\theta_{23}=\pi/4,\delta=\pi/2,3\pi/2$)\cite{H1,H2,H3,H4,H5} would  undergo a serious experimental test.\\

 Despite having a decent theoretical understanding at least on the leading order neutrino mixing\cite{Alt} and other testable parameters such as $\delta$\cite{pet}, the paradigm of residual symmetry\cite{Lam,dicus,he} turns out to be one of the  most economical and promising approach since one really does not require  the values of the elements of the light neutrino mass matrix $M_\nu$ to predict the mixing parameters\cite{Lam}. For a given neutrino Majorana mass matrix $M_\nu$ and a  representation of the residual symmetries in the lepton flavour space $\mathcal{G}_a$, a horizontal invariance $(\mathcal{G}_a)^T M_\nu\mathcal{G}_a=M_\nu$ ($a=1,2,3$) together with the diagonalization condition $U^TM_\nu U={\rm diag}(m_1,m_2,m_3)$ imply
\bea
\hspace{2cm}\mathcal{G}_{a}=U d_{a}U^\dag \hspace{2mm}{\rm with}\hspace{2mm}(d_a)_{ij}=\pm \delta_{ij},\label{eq1}
\eea
 where  $U$ is a unitary matrix that diagonalizes $M_\nu$ having nondegenerate eigenvalues. Thus the columns of the mixing matrix are simply the eigenvectors of horizontal symmetry matrix $\mathcal{G}_a$ with eigenvalues $\pm 1$. It can be shown that out of eight possible $d_a$, only two are independent\cite{Lam} which lead to a closure property $\mathcal{G}_a \mathcal{G}_b=\mathcal{G}_c$ with $a\neq b\neq c$. Since each $\mathcal{G}$ generates a $\mathbb{Z}_2$ symmetry, the entire neutrino mixing could be interpreted as a consequence of a residual $\mathbb{Z}_2\times\mathbb{Z}_2$ symmetry.
It also follows from Eq.\ref{eq1} that $\mathcal{G}^2=I,\hspace{2mm}\det \mathcal{G}=\pm 1$. One can restrict to, without any loss of generality $\det \mathcal{G}=+1$ leading to two independent choices of $d$: $d_1={\rm diag}\hspace{1mm}(1,-1,-1)$ and $d_2={\rm diag}\hspace{1mm}(-1,1,-1)$ with $d_3=d_1d_2$. The choice $\det \mathcal{G}=-1$ would then be a trivial option with all the $d$ matrices same upto an overall minus sign. Thus, given a neutrino mixing matrix $U$ and $d_{1,2}$, one can construct the corresponding $\mathbb{Z}_2$ generators $\mathcal{G}_{1,2}$ using Eq.\ref{eq1}. For  example, within the PDG convention\cite{pdg}, a leading order mixing matrix $U_0^{\mu\tau}$ could be used along with $d_3$ to reconstruct the well known $\mu\tau$ interchange symmetry $\mathcal{G}_3^{\mu\tau}$\cite{mutau} matrix:
\bea
U_0^{\mu\tau}=\begin{pmatrix}p & x & 0\\-\frac{x}{\sqrt{2}} & \frac{p}{\sqrt{2}} & \frac{1}{\sqrt{2}}\\ \frac{x}{\sqrt{2}} & -\frac{p}{\sqrt{2}} & \frac{1}{\sqrt{2}}\end{pmatrix}\xRightarrow[\text{(\ref{eq1})}]{\text{ $d_3$}}\mathcal{G}_3^{\mu\tau}=\begin{pmatrix}
-1&0&0\\0&0&1\\0&1&0
\end{pmatrix},\label{eq2}
\eea  
where, $p\equiv\cos\theta_{12}^\circ$ and $x\equiv\sin\theta_{12}^\circ$ with $\theta_{12}^\circ$ being the solar mixing angle. Using Eq.\ref{eq1} and  $d_{1,2}$, the other two matrices, i.e., $\mathcal{G}^{\mu\tau}_{1,2}$ can easily be constructed. We would like to call them the associate $\mu\tau$ symmetries. Since a nonzero value ($\sim9^\circ$) for the reactor mixing angle has been confirmed at more than $5.2\sigma$\cite{An:2015rpe}, $\mathcal{G}_3^{\mu\tau}$ invariance must be broken in the neutrino mass term. However, due to the presence of  $\mathcal{G}^{\mu\tau}_1$ ($\mathcal{G}^{\mu\tau}_2$) and the corresponding $d_{1}$ ($d_{2}$) matrix, one can simply `rotate' $U_0^{\mu\tau}$ in 2-3 (1-3) plane to obtain a nonzero reactor mixing angle $\theta_{13}$ and hence the phase $\delta$. It is to be noted, that these rotations are very natural possibilities due to the degenerate eigenvalues of $\mathcal{G}^{\mu\tau}_{1,2}$ matrices. For any light neutrino mass matrix that enjoys such an invariance, the authors of Ref.\cite{dicus} showed that the  phase $\delta$ turns out to be
\bea
\hspace{9mm}\cos\delta=\frac{(s_{23}^2-c_{23}^2)(s_{12}^2-c_{12}^2s_{13}^2)}{4c_{12}s_{12}c_{23}s_{23}s_{13}}\hspace{1cm}{\rm for}\hspace{2mm}\mathcal{G}_1^{\mu\tau},\label{g1d}\\
\hspace{9mm}\cos\delta=\frac{(c_{23}^2-s_{23}^2)(c_{12}^2-s_{12}^2s_{13}^2)}{4c_{12}s_{12}c_{23}s_{23}s_{13}}\hspace{1cm}{\rm for}\hspace{2mm}\mathcal{G}_2^{\mu\tau},\label{g2d}
\eea
where $\theta_{23},\theta_{12},\theta_{13}$ are the atmospheric, solar and reactor mixing angles respectively. For a given $3\sigma$ range\cite{nufit} of $\theta_{12}$ and $\theta_{13}$, the relations in Eq.\ref{g1d} and Eq.\ref{g2d}  predict a simultaneous maximality ($\delta=\pi/2$ or $3\pi/2$ and $\theta_{23}=\pi/4)$. Keeping in mind that still there are no definite statements regarding the values of $\delta$ and $\theta_{23}$,  we propose a generalization of the associate $\mu\tau$ interchange symmetries and we refer to them as associate $\mu\tau$ mixing symmetries. The mixing symmetry could relax the simultaneous maximality of $\delta$ and $\theta_{23}$, i.e., unlike the prediction of exact $\mu\tau$ interchange (cf. Eq.\ref{g1d}), in this scenario, nonmaximal value of $\delta$ ($\theta_{23}$) is allowed for a maximal value of   $\theta_{23}$ ($\delta$). General $\mu\tau$ symmetry is  basically a `mixing' between $\mu$ and $\tau$ neutrino flavours unlike the conventional $\mu\tau$ `interchange'. Similar to the $\mathcal{G}_3^{\mu\tau}$ generator (cf. Eq.\ref{eq2}), we can derive the same (we designate it as $\mathcal{G}_3^{g\mu\tau}$) for the $\mu\tau$ mixing, starting   from the leading order mixing matrix  $U_0^{g\mu\tau}$ as
\bea
\hspace{2cm} U_0^{g\mu\tau}=\begin{pmatrix}p & x & 0\\-xq & pq & y\\xy & -py & q\end{pmatrix},\label{Umix}
\eea
where $y=\sin\theta_g$ and $q=\cos\theta_g$ with $\theta_g$ being the $\mu\tau$ mixing parameter and using Eq.\ref{eq1}. Thus $\mathcal{G}_3^{g\mu\tau}$ could be constructed  as
\bea
\hspace{1.5cm}\mathcal{G}_3^{g\mu\tau}=\begin{pmatrix}
-1&0&0\\
0&-\cos2\theta_g&\sin 2\theta_g\\0&\sin2\theta_g&\cos2\theta_g
\end{pmatrix}.\label{mixG}
\eea 
Note that for $\theta_g=\pi/4$, we recover the usual $\mu\tau$ interchange symmetry $\mathcal{G}_3^{\mu\tau}$ (cf. Eq.\ref{eq2}). For an elaborate discussion regarding the high energy flavour models that break down to low energy residual symmetries like $\mathcal{G}_3^{g\mu\tau}$, we refer to \cite{Grimus:2004rj,Grimus:2004cc}. Similar to \cite{dicus}, in our proposal also, due to the vanishing $\theta_{13}$, we opt for the predictions of the associate $\mu\tau$ mixing symmetries $\mathcal{G}_{1,2}^{g\mu\tau}$  assuming that the $\mathcal{G}_3^{g\mu\tau}$ is broken. It is now trivial to anticipate, that the parameter $\theta_g\neq \pi/4$ is entirely responsible for  $\theta_{23}$ and $\delta$ not being simultaneously maximal. Having set up all the necessary prerequisites,  we opt for a three step presentation of this paper. In the first step, we primarily obtain $\delta$ as $\delta\equiv f(\theta_{23},\theta_{12},\theta_{13},\theta_g)$ (Eq.\ref{g1u} \& \ref{g2u})
for both the associate $\mu\tau$ mixing symmetries in a model independent way. Then we present a very general numerical analysis. For example,  for the given maximality of $\theta_{23}$ ($\delta$) we try to show how far  $\theta_g$ could deviate from $\pi/4$ for the allowed nonmaximal value of $\delta$ ($\theta_{23}$). We find that the deviation (measured by a parameter $\theta_d$ which is related to $\theta_g$ as $\theta_g=\pi/4\pm \theta_d$) cannot be very large, in particular for $\mathcal{G}_2^{g\mu\tau}$, the deviation is significantly small. To compare our results with exact $\mu\tau$ interchange, we then present a distribution of $\cos\delta$ for a small value of $\theta_d$ taking into account a Gaussian distribution  for the other mixing angles. In the second step, we discuss how the parameter $\theta_g$ could be related to a realistic model parameter in neutrino mass models such as softly broken $D_4$\cite{Grimus:2004cc}, Scaling Ansatz \cite{Mohapatra:2006xy,joshi,CP1,CP2,CP3,CP4,CP5}, four texture zeros in neutrino Dirac mass matrix within Type-I seesaw\cite{Branco:2007nb} etc. In the third step, we show how the associate mixing symmetries  $\mathcal{G}_{1,2}^{g\mu\tau}$ could be a consequence of two simultaneous CP transformations\cite{CPt1,CPt2,mu1} in  the neutrino mass terms. We then derive  novel correlations between $\delta$ and the mixing angles $\theta_{ij}$ in this class of models. Finally, from the perspective of  leptogenesis,   we show this CP extended $\mu\tau$ mixing symmetry is more interesting than the CP extended $\mu\tau$ interchange which has been a subject of recent interest in neutrino mass models\cite{mu2,mu3,mu4,mu5,mu6,mu7,mu8,mu9,mu10,mu11,Samanta:2015oqa,mu12,mu13}. In particular, we focus on the `three flavour regime' of leptogenesis\cite{lep1,lep2,lep3} and show,  unlike the CP extended $\mu\tau$ interchange, a resonant leptogenesis\cite{reso} is possible in our scheme. and {\it in this framework, a nonzero value of baryon asymmetry always requires nonmaximality in $\theta_{23}$}.  We also demonstrate that the CP extended $\mu\tau$ mixing symmetry is a novel and nice example that shows the importance of the off-diagonal terms of the flavour coupling matrix\cite{lep4,lep5,lep6,lep7} which have usually been neglected in the computation of leptogenesis, particularly, in the models with flavour symmetries. \\

The rest of the paper is organized as follows. Sec.\ref{s2} and its various subsections deal with the explicit theoretical formalism to derive the model independent constraints and some pictorial representations of the sensitivity of the parameters $\delta$ and $\theta_{23}$ with the newly introduced parameter $\theta_g$ that generalizes $\mu\tau$ interchange to $\mu\tau$ mixing.  We then compare our results with the exact $\mu\tau$ interchange symmetry and discuss the significance of the parameter $\theta_d$ in neutrino mass models such as Scaling Ansatz. In Sec.\ref{s3}  we demonstrate the CP extended $\mu\tau$ mixing and its consequences. In Sec.\ref{s4} we present a qualitative as well as a quantitative description of leptogenesis within the framework of CP extended $\mu\tau$ mixing. Finally, we conclude our work  in Sec.\ref{s5}.
 \section{Model independent correlations in $\mu\tau$ mixing symmetry}\label{s2}
 In this section, for both the associate mixing symmetries, first we derive  analytical correlations among the Dirac CP violating phase, mixing angles and the proposed mixing parameter $\theta_g$. Then we try to show the compatibility of the scenario with recent neutrino oscillation data\cite{nufit}. A systematic analysis is given in what follows.
\subsection{Consequences of $\mathcal{G}_1^{g\mu\tau}$ invariance} 
As already introduced in the previous section, the matrix $d_1={\rm diag}(1,-1,-1)$ has two degenerate entries. Therefore, given the symmetry $\mathcal{G}_1^{g\mu\tau}$ and the diagonalization condition in Eq.\ref{eq1}, the second and third columns of the mixing matrix $U_0^{g\mu\tau}$ are not unique. It could be rotated in the 2-3 plane due to the aforementioned two-fold degeneracy. This is intriguing because the phenomenological requirement of having a nonvanishing reactor mixing angle finds a natural symmetry justification. With the choice of a most general unitary rotation matrix $U_\theta^{23}$\cite{dicus} in the 2-3 plane 
\begin{equation}
\hspace{1cm}U_\theta^{23}=\begin{pmatrix}
1 & 0 & 0\\0 & c_{\theta} & s_{\theta}e^{i\gamma}\\0 & -s_{\theta}e^{-i\gamma} & c_{\theta}
\end{pmatrix}P_\phi,\label{rot1}
\end{equation}
 where $P_{\phi}={\rm diag}(e^{i\phi_1},e^{i\phi_2},e^{i\phi_3})$,
 a phenomenologically consistent PMNS matrix $U=U_0^{g\mu\tau}U_\theta^{23}$ is obtained as \begin{equation}
U=\begin{pmatrix} p & xc_{\theta} & xs_{\theta}e^{i\gamma}\\-xq & (pqc_{\theta}-ys_{\theta}e^{-i\gamma}) & (yc_{\theta}+pqs_{\theta}e^{i\gamma})\\xy & -(pyc_{\theta}+qs_{\theta}e^{-i\gamma}) & (qc_{\theta}-pys_{\theta}e^{i\gamma})\end{pmatrix}P_\phi.\label{xx1}
\end{equation}
We now compare Eq.\ref{xx1} to the PMNS matrix which is parametrized according to the PDG convention\cite{pdg} as\\
\begin{equation}
U_{\rm PMNS}=P_{\chi}\begin{pmatrix}
c_{12}c_{13} & s_{12}c_{13} & s_{13}e^{-i\delta}\\-s_{12}c_{23}-c_{12}s_{23}s_{13}e^{i\delta} & c_{12}c_{23}-s_{12}s_{23}s_{13}e^{i\delta} & c_{13}s_{23}\\s_{12}s_{23}-c_{12}c_{23}s_{13}e^{i\delta} & -c_{12}s_{23}-s_{12}c_{23}s_{13}e^{i\delta} & c_{13}c_{23}\end{pmatrix}P_{M}, \label{k11}
\end{equation} 
where $P_{\chi}={\rm diag}\hspace{1mm}(e^{i\chi_1},e^{i\chi_2},e^{i\chi_3})$ is an unphysical phase matrix and $P_{M}={\rm diag}\hspace{1mm}(1,e^{i\frac{\alpha}{2}},e^{i\frac{\beta}{2}})$ represents the Majorana phase matrix.\\

Comparing the (11), (12) and (13) element of Eq.\ref{xx1} and Eq.\ref{k11}, we find
\begin{flalign}
c_{12}c_{13}=p,\hspace{0.2cm} \chi_1=\phi_1,\label{a1}\\
 s_{12}c_{13}=xc_{\theta}, \hspace{0.2cm}\frac{\alpha}{2}+\chi_1=\phi_2,\label{a2}\\
\hspace{1cm} s_{13}=xs_{\theta},\hspace{0.2cm} \chi_1-\delta+\frac{\beta}{2}-\gamma=\phi_3.\label{a3}
\end{flalign}
Equating the quantity $|U_{21}|^2-|U_{31}|^2$ of Eq.\ref{xx1} with the same of Eq.\ref{k11} and using Eq.\ref{a1}-Eq.\ref{a3}, we obtain
\begin{equation}\cos\delta=\frac{(s_{23}^2-c_{23}^2)(s_{12}^2-c_{12}^2s_{13}^2)+\cos2\theta_g(s_{13}^2+c_{13}^2s_{12}^2)}{4c_{12}s_{12}c_{23}s_{23}s_{13}},\label{g1u}
\end{equation} where we have re-expressed $y,q$ in terms of $\theta_g$. As expected, for $\theta_g=\pi/4$, Eq.\ref{g1u} reduces to Eq.\ref{g1d} which is the prediction of $\mathcal{G}_1^{\mu\tau}$. For numerical purpose, it is convenient to parametrize $\theta_g$ as $\theta_g=\pi/4+\theta_d$. With this parametrization, it would be easier to realize the variation of the observables with respect to $\theta_d$ which is a measure of the deviation from the usual $\mu\tau$ interchange symmetry. Introduction of the mixing parameter $\theta_g$ now enables us to explore various interesting aspects of Eq.\ref{g1u}. For example, if we set the atmospheric mixing angle $\theta_{23}$ to be maximal, the deviation of $\cos\delta$ from its maximal value can be tracked with $\theta_d$  from 
\bea
\hspace{1cm}\cos\delta=\frac{\cos2(\pi/4+\theta_d)(s_{13}^2+c_{13}^2s_{12}^2)}{2c_{12}s_{12}s_{13}}.
\eea
\begin{figure}[H]
\includegraphics[scale=.4]{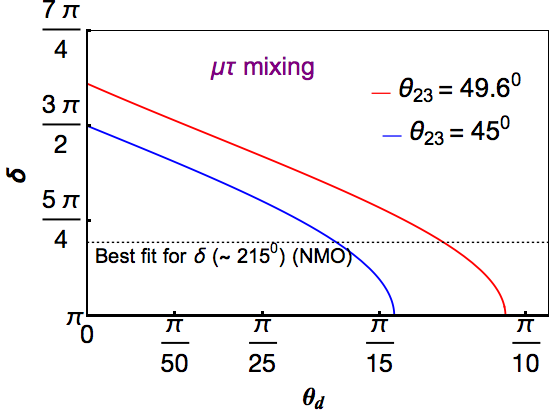} \includegraphics[scale=.4]{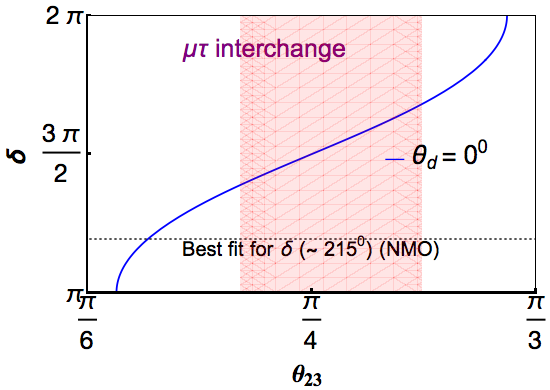}
\caption{For $\mathcal{G}_1^{g\mu\tau}$(left): Variation of $\delta$ with $\theta_d$, where the latter is a measure of deviation from $\mu\tau$ interchange symmetry or the strength of $\mu\tau$ mixing. Here, 2$\pi-\delta$ is also an allowed  solution for the same values of $\theta_d$. For $\mathcal{G}_1^{\mu\tau}$(right):  Variation of $\delta$ with $\theta_{23}$. These plots are generated using the best-fit values of $\theta_{13}$ and $\theta_{12}$ for Normal Mass Ordering\cite{nufit}.}\label{fig1}
\end{figure}  
Similarly, for a maximal Dirac CP violation (a general expression of $\theta_{23}$ is given in the appendix), one obtains
\bea
\cos 2\theta_{23}=\frac{\cos 2(\pi/4+\theta_d)(s_{13}^2+c_{13}^2s_{12}^2) }{s_{12}^2-c_{12}^2s_{13}^2}.
\eea
For the best-fit values of $\theta_{12}$ and $\theta_{13}$ (TABLE \ref{t1}), we present a variation of $\delta$ with $\theta_d$ (left) for $\mu\tau$ mixing and $\delta$ with $\theta_{23}$ for $\mu\tau$ interchange\footnote{From now on when we address predictions  of $\mu\tau$ mixing or $\mu\tau$ interchange, it could be assumed that we are implying the predictions for the associate symmetries.  } (right)  in Fig.\ref{fig1}. \\

\begin{figure}[H]
\includegraphics[scale=.55]{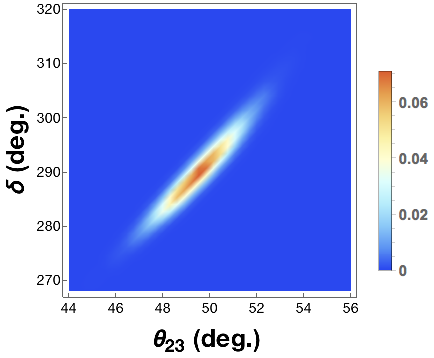} \includegraphics[scale=.55]{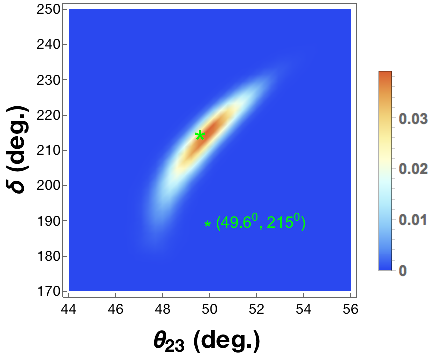}
\caption{For $\mathcal{G}_1^{\mu\tau}$ (left): Probability distribution of $\delta$ with $\theta_{23}$. For $\mathcal{G}_1^{g\mu\tau}$ (right): The same plot but for $\theta_d=\pi/12$. Here we have used Gaussian distribution for each of the mixing angles.}\label{fig2}
\end{figure}

It is evident from the first figure (left) in Fig.\ref{fig1}, that the robust prediction of simultaneous maximality coming from  $\mu\tau$ interchange  (cf. Eq.\ref{g1d}) has now  been relaxed (represented by the blue line for $\theta_{23}=45^0$) since $\theta_d$ can be different from zero. In fact, one can see that the deviation of $\delta$ from its maximal value is very much sensitive to $\theta_d$, e.g., a deviation of the former from $3\pi/2$ to $5\pi/4$  only requires a value $\approx\pi/20$ for the latter. The red line represents the variation of $\delta$ for the current best-fit value $\theta_{23}=49.6^0$ for NMO. As one can see, in the $\mu\tau$ interchange limit ($\theta_d=0$) best-fit of $\theta_{23}$ is not consistent with the current best-fit of $\delta=215^0$ (represented by the horizontal black dotted line).  However,  in the proposed $\mu\tau$ mixing scheme,  one can fit the best-fits simultaneously just by tuning a single parameter $\theta_d ~(\sim \pi/12)$ as shown by the red line. The figure in the right hand side shows a variation of $\delta$ with $\theta_{23}$ for the exact $\mu\tau$ interchange symmetry. It is interesting to notice, that even within the  $3\sigma$ range of $\theta_{23}$, one cannot reconcile the best-fit value of $\delta$. Thus from the viewpoint of current experimental results, the proposed $\mathcal{G}_1^{g\mu\tau}$ is a more admissible symmetry than the $\mathcal{G}_1^{\mu\tau}$. This can also be realized more clearly from the Fig.\ref{fig2} where we present a statistical comparison between the predictions of $\mu\tau$ interchange ($\mathcal{G}_1^{\mu\tau}$) and $\mu\tau$ mixing ($\mathcal{G}_1^{g\mu\tau}$). The probability density plot\footnote{These are basically  density-histogram plots. As one goes from the blue region to the red one, the number density of points increases, as shown in the bar-legends.} in the left hand side of Fig.\ref{fig2} shows, for the $\mathcal{G}_1^{\mu\tau}$, most probable values of $\delta$ lie within a region centred approximately around $\delta\sim 290^0$ which is far away (tension is $\sim 2\sigma$) from the best-fit $215^0$ for NMO.  
On the other hand, as also explained earlier, for the best-fit of $\theta_{23}$, the most probable values of $\delta$ could be reconciled with the best-fit $215^0$ for $\theta_{d}=\pi/12$. As shown in right panel of Fig.\ref{fig2}, the entire pattern (shown in the  left  side) has shifted near the best-fit shown by the green `$*$'.  Before discussing the consequences of $\mathcal{G}_2^{g\mu\tau}$, let us us quickly address some important points. First of all, when we say the  $\mu\tau$ interchange ($\mathcal{G}_1^{\mu\tau}$) is disfavoured, we always mean a NMO. As one can see from Fig.\ref{fig2} (left hand side), best-fit of $\delta~(=284^0)$ for IMO could be well reconciled within $1\sigma$ of $\theta_{23}$. However, as mentioned in the introduction, that the IMO seems to be disfavoured now by the current experimental data. One might also wonder how the mixing parameter $\theta_{g}=\pi/4\pm\theta_d$ could be realized in a realistic neutrino mass model. Because so far it appears to be a model independent tuning parameter, except the mention to Ref.\cite{Grimus:2004rj,Grimus:2004cc} in the introduction. In Sec.\ref{s2.3}, we shall briefly discuss some models regarding the relation of $\theta_g$ with the model parameters and show indeed there is a large class of models that exhibit $\mu\tau$ mixing at low energy.

\subsection{Consequences of $\mathcal{G}_2^{g\mu\tau}$ invariance}
 In this case, a rotation in 1-3 plane is possible due to the degeneracy in $d_2={\rm}~(-1,1,-1)$ matrix. By choosing a most general unitary rotation matrix $U_\theta^{13}$\cite{dicus} as 
\begin{equation} 
 \hspace{1cm}U_\theta^{13}=\begin{pmatrix} c_{\theta} & 0 & s_{\theta}e^{i\gamma}\\0 & 1 & 0\\-s_{\theta}e^{-i\gamma} & 0 & c_{\theta} \end{pmatrix} P_{\phi}\label{rot2}
\end{equation}
 we construct the phenomenologically viable PMNS matrix  $U=U_0^{g\mu\tau}U_\theta^{13}$ which is given by
 \begin{equation}
 U=\begin{pmatrix} pc_{\theta} & x & ps_{\theta}e^{i\gamma}\\-(xqc_{\theta}+ys_{\theta}e^{-i\gamma}) & pq & (yc_{\theta}-xqs_{\theta}e^{i\gamma})\\(xyc_{\theta}-qs_{\theta}e^{-i\gamma}) & -py & (qc_{\theta}+xys_{\theta}e^{i\gamma})\end{pmatrix} P_{\phi}.\label{k4}
 \end{equation}
 Following a similar procedure  demonstrated previously for $\mathcal{G}_1^{g\mu\tau}$, we compare Eq.\ref{k4} with the PMNS matrix of Eq.\ref{k11} and equate the quantity $|U_{22}|^2-|U_{32}|^2$ of both the matrices. This results in
\begin{equation}
\cos\delta=\frac{(c_{23}^2-s_{23}^2)(c_{12}^2-s_{12}^2s_{13}^2)-\cos 2\theta_g(s_{13}^2+c_{13}^2c_{12}^2)}{4c_{12}s_{12}c_{23}s_{23}s_{13}}.\label{g2u}
\end{equation}
Thus for $\theta_g=\pi/4$, we recover Eq.\ref{g2d}. In this case,  parametrizing \footnote{Unlike the case of $\mathcal{G}_1^{g\mu\tau}$, here we parametrize $\theta_g$ as $\theta_g=\pi/4-\theta_d$ instead of $\theta_g=\pi/4+\theta_d$, just to show the variation of $\delta$ with the positive values of $\theta_d$.}  $\theta_g=\pi/4-\theta_d$, one can derive the simple correlations of $\delta$ and $\theta_{23}$ with $\theta_g$ as
\begin{equation}
\cos\delta=\frac{-\cos 2(\pi/4-\theta_d)(s_{13}^2+c_{13}^2c_{12}^2)}{2c_{12}s_{12}s_{13}},
\end{equation}
\bea
\cos 2\theta_{23}=\frac{\cos 2(\pi/4-\theta_d)(s_{13}^2+c_{13}^2c_{12}^2) }{c_{12}^2-s_{12}^2s_{13}^2}
\eea
for a maximal atmospheric angle and maximal Dirac CP violation respectively.
\begin{figure}[H]
\includegraphics[scale=.4]{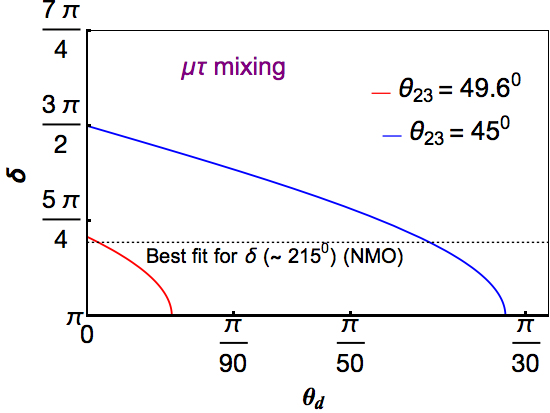} \includegraphics[scale=.4]{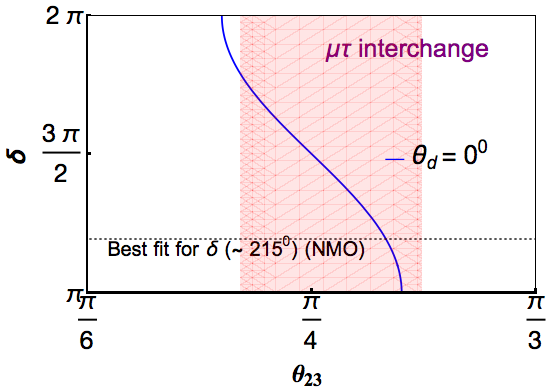}
\caption{For $\mathcal{G}_2^{g\mu\tau}$(left): Variation of $\delta$ with $\theta_d$, where the latter is a measure of deviation from $\mu\tau$ interchange symmetry or the strength of $\mu\tau$ mixing. Here, 2$\pi-\delta$ is also an allowed  solution for the same values of $\theta_d$. For $\mathcal{G}_2^{\mu\tau}$(right):  Variation of $\delta$ with $\theta_{23}$. These plots are generated using the best-fit values of $\theta_{13}$ and $\theta_{12}$ for Normal mass ordering\cite{nufit}.}\label{fig3}
\end{figure}  

\begin{figure}[H]
\includegraphics[scale=.55]{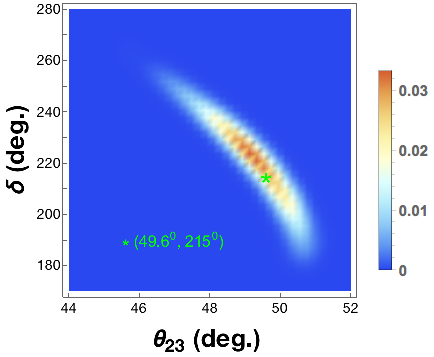} \includegraphics[scale=.55]{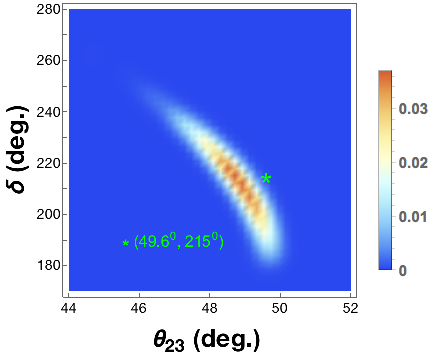}
\caption{For $\mathcal{G}_2^{\mu\tau}$ (left): Probability distribution of $\delta$ with $\theta_{23}$. For $\mathcal{G}_2^{g\mu\tau}$ (right): The same plot but for $\theta_d=\pi/180$. Here we have used Gaussian distribution for each of the mixing angles.}\label{fig4}
\end{figure}  
In Fig.\ref{fig3} (left figure), the red line represents the variation of $\delta$ with $\theta_d$ for the best-fit of $\theta_{23}=49.6^0$. We find a remarkable `coincidence' of Eq.\ref{g2d} with the present data on $\theta_{23}$ and $\delta$. It is evident from the plot in the left panel of Fig.\ref{fig3} (also in  Fig.\ref{fig4}), one needs really a tiny departure (numerically, $\theta_d$ less than $0.5^0$) from exact $\mu\tau$ interchange to fit the most probable values of $\delta$  simultaneously with the best-fit of $\delta$ and $\theta_{23}$. When we opt for a larger departure from $\mu\tau$ interchange, even with $\theta_d=1^0$, the most probable values of $\delta$ start to move towards CP conserving values  as shown in the right panel of Fig.\ref{fig4}. Thus,  as far as the current data on $\delta$ and $\theta_{23}$ is concerned, undoubtedly, $\mu\tau$ interchange (here $\mathcal{G}_2^{\mu\tau}$) is a better symmetry to explain the data than the proposed $\mu\tau$ mixing (here $\mathcal{G}_2^{g\mu\tau}$). However, compared to the previously released data\cite{old}, the new data\cite{nufit} shows a tendency to move towards the CP conserving values mainly driven by NO$\nu$A anti-neutrino appearance channel\cite{nufit}. If this trend continues, one has to think beyond $\mu\tau$ interchange symmetry. In that case, the $\mu\tau$ mixing (as shown in the  right panel of  Fig.\ref{fig4}) could be a good option to explain the data.\\

So we  conclude this section with the remark that, to explain the present data,  the proposed $\mathcal{G}_1^{g\mu\tau}$ (mixing) symmetry has an edge over the $\mathcal{G}_1^{\mu\tau}$ (interchange), whereas the symmetries which are in the class of $\mathcal{G}_2$, to explain the present data, the interchange scenario is  better than the proposed mixing scenario. 

\subsection{$\mu\tau$ mixing in some neutrino mass models} \label{s2.3}
In this section, we  discuss some examples of the low energy residual $\mu\tau$ symmetry so that the parameter $\theta_g$ could be connected to the model parameter(s). Indeed there is a large class of models that belong to the mixing category. For example the authors of Ref.\cite{Grimus:2004rj,Grimus:2004cc} derive the mixing from explicit symmetry group $D_4$ where the $\mu\tau$ mixing parameter $\theta_g$ could be related to the model parameter as
\bea
\cos 2 \theta_g \sim -\frac{\mu_{\rm soft}}{M},\label{soft}
\eea
where  $\mu_{\rm soft}$  is a soft breaking term in the $D_4$ model and $M$ is the mass scale of the  RH neutrinos needed to obtain Type-I seesaw light neutrino masses. The model predicts the same leading order mixing matrix as shown in Eq.\ref{Umix}. Now to generate a nonzero $\theta_{13}$ we can further add soft breaking terms to the model. However, notice that, here the introduction of new breaking terms corresponds to the rotation of the leading order mixing matrix (cf. Eq.\ref{rot1}) and the final prediction (cf. Eq.\ref{g1u}) is independent of the rotation angle. Since in general, generation of nonzero $\theta_{13}$ requires small breaking terms,  Eq.\ref{soft} still holds at the leading order. But unlike the $D_4$ model, $\cos 2\theta_g$ is not directly related only to $\theta_{23}$ but it connects   $\theta_{23}$ and $\delta$ with the correlations shown in Eq.\ref{g1u} or Eq.\ref{g2u}.  The authors of Ref.\cite{Grimus:2004rj,Grimus:2004cc} conclude, that to test a sizeable deviation of physical parameters such as $\theta_{23}$ one needs the scale of $\mu_{\rm soft}$ of same order as the mass scale of the RH neutrinos requiring a large deviation from $\mu\tau$ interchange ($\cos 2\theta_g\simeq 1$).  However, as we have shown  in the previous section, even with a small departure from the interchange symmetry, one can test the parameters $\delta$ and $\theta_{23}$ with the correlations derived in this paper. Thus the scale of $\mu_{\rm soft}$ and $M$ need not be the same. In fact one needs  much smaller scale for $\mu_{\rm soft}$ than the scale $M$. Models with Scaling Ansatz or Simple Real Scaling (SRS) originally proposed in Ref.\cite{Mohapatra:2006xy} and then analysed at length in Ref.\cite{joshi,CP1,CP2,CP3,CP4,CP5,Samanta:2015hxa}, also belong to the $\mu\tau$ mixing category. As derived in Ref.\cite{CP1}, the residual symmetry for SRS is given by 
\bea
G_3^{k}=\begin{pmatrix}
-1&0&0\\0&(1-k^2)(1+k^2)^{-1}&2k(1+k^2)^{-1}\\0&2k(1+k^2)^{-1}&-(1-k^2)(1+k^2)^{-1}
\end{pmatrix},
\eea
where `$k$' is the scale parameter of the model that scales, for example, one row of $M_\nu$ with the other row or one column with another column\cite{Mohapatra:2006xy}.
Thus the scale parameter of the model could be constrained simply by the relation 
\bea
k=\frac{1\pm \cos 2\theta_g}{\sin 2\theta_g}
\eea
along with Eq.\ref{g1u} or Eq.\ref{g2u}. One can also constrain the  parameters of models like four zero textures (in charge lepton flavour basis) within Type-I seesaw\cite{Branco:2007nb}, so called discrete Dark Matter (DM) models\cite{dis1,dis2} as well as the models with global $U(1)$ symmetries\cite{u11,u12,u13,u14} which at the leading order, show up a $\mu\tau$ mixing scheme at  low energy.  However, we would like to stress that at this stage, where the precise values of $\delta$ and $\theta_{23}$ are yet to be measured, if the $\mu\tau$ mixing parameter can be constrained a priori by some other constraints, (e.g., the mass-squared differences, the mixing angles other than $\theta_{23}$ or some cosmological phenomenon such as leptogenesis etc.) then Eq.\ref{g1u} or Eq.\ref{g2u} could be used to predict $\delta$ or $\theta_{23}$. So the models with $\mu\tau$ mixing symmetry and lesser number of parameters (such as Scaling Ansatz+ texture zeros\cite{CP5}) are most welcome.
\section{CP extension of the  $\mu\tau$ mixing symmetry}\label{s3}
\subsection{CP symmetry in a general light neutrino mass term}
So far the discussion was quite general. In this section we want to explore some special class of $\mu\tau$ mixing. Since in general, only flavour symmetries are not sufficient to predict the CP violating phases, a lot of effort has been devoted in past few years to ameliorate flavour symmetries with CP symmetries\cite{mu13} by demanding the invariance of the neutrino mass term with the field transformation 
 \begin{equation} 
 \nu_{L\ell }\to i(\mathcal{G}_a)_{\ell m}\gamma^0\nu^C_{Lm}~~(a=1,2,3).\label{CPtr}
 \end{equation} 
 Though one always has to respect the `consistency condition' to have a combined theory of flavour and CP\cite{mu2,mu3}. The consistency condition can be written as
\bea
X_r\rho_r^*(g)X_r^{-1}=\rho_r(g^\prime),\label{const}
\eea
where  $X_r$ is a unitary matrix representing CP symmetry  which acts on a generic multiplet $\varphi$  as
\bea
\varphi(x)\xrightarrow{\text{CP}} X_r\varphi^*(x^\prime)
\eea
with $x^\prime=(t,-{\bf x})$ and $\rho_r(g)$, $\rho_r(g^\prime)$ are the representations for the elements $g$, $g^\prime$ of the flavour group in an irreducible representation ${\bf r}$. Eq.\ref{CPtr} leads to the complex invariance
 \begin{equation} 
 \mathcal{G}_{a}^TM_\nu \mathcal{G}_{a}=M_\nu^*.
 \end{equation} 
  Now at low energy, among the three residual $\mathbb{Z}_2$ generators, if two of them, say,
$\mathcal{G}_2$ and $\mathcal{G}_3$ correspond to the complex invariances
\begin{equation} 
 \mathcal{G}_{2}^TM_\nu \mathcal{G}_{2}=M_\nu^*, \mathcal{G}_{3}^TM_\nu \mathcal{G}_{3}=M_\nu^*,\label{ci}
 \end{equation}
 the remaining one, i.e., $\mathcal{G}_{1}$ automatically satisfies a real invariance\cite{mu1}
  \begin{equation} 
 \mathcal{G}_{1}^TM_\nu \mathcal{G}_{1}=M_\nu.
 \end{equation} 
 It is trivial now to show, that Eq.\ref{const} is satisfied. Since in this case, the CP and flavour symmetry generators are basically $\mathbb{Z}_2$ generators, Eq.\ref{const} would imply 
 \bea
 \mathcal{G}_2 \mathcal{G}_1 \mathcal{G}_2^{-1}=\mathcal{G}_1,\\
 \mathcal{G}_3 \mathcal{G}_1 \mathcal{G}_3^{-1}=\mathcal{G}_1.
 \eea
 Since by construction, $\mathcal{G}_a\mathcal{G}_b=\mathcal{G}_b\mathcal{G}_a=\mathcal{G}_c $ for $a\neq b \neq c$, the left and right hand sides of the above equations are consistent.  Similarly, one obtains a real invariance for $\mathcal{G}_2$, for the simultaneous complex invariances for $\mathcal{G}_1$ and $\mathcal{G}_3$. However, note that, if we demand the complex invariances for $\mathcal{G}_1$ and $\mathcal{G}_2$, we obtain a real invariance for $\mathcal{G}_3$ which is not acceptable, since, that will correspond to a vanishing $\theta_{13}$. Let us now turn into the computation of the Dirac CP phase for both the acceptable real invariances, i.e., $ (\mathcal{G}_{1}^{g\mu\tau})^TM_\nu \mathcal{G}_{1}^{g\mu\tau}=M_\nu$ and $(\mathcal{G}_{2}^{g\mu\tau})^TM_\nu \mathcal{G}_{2}^{g\mu\tau}=M_\nu$. For  both the cases, the second of \eqref{ci} leads to\cite{Chen:2015siy,Sinha:2018xof}
 \begin{equation}
\sin\delta=\pm\frac{\sin2\theta_g}{\sin2\theta_{23}}.\label{g3c}
\end{equation}
 Eliminating $\theta_g$ from \eqref{g1u} and \eqref{g3c} for $\mathcal{G}_1^{g\mu\tau}$ whereas doing the same from \eqref{g2u} and \eqref{g3c} for $\mathcal{G}_2^{g\mu\tau}$ we obtain a generic expression for $\cos\delta$ as
\bea
\cos\delta =\frac{A_iB \pm \sqrt{A_i^2B^2-(B^2-C_i^2\sin^2 2\theta_{23})(A_i^2-C_i^2\cos^2 2\theta_{23})}}{(B^2-C_i^2\sin^2 2\theta_{23})},\label{cd}
\eea
where $i=1,~2$ corresponds $\mathcal{G}_i^{g\mu\tau}$ symmetries.
The parameters $A_i,B$ and $C_i$ are the functions of the mixing angles $\theta_{ij}$ with the explicit expressions
\bea
A_1=(s_{23}^2-c_{23}^2)(s_{12}^2-c_{12}^2s_{13}^2), \hspace{3mm}A_2=(c_{23}^2-s_{23}^2)(c_{12}^2-s_{12}^2s_{13}^2),\\
C_1=(s_{13}^2+c_{13}^2s_{12}^2),\hspace{5mm} C_2=-(s_{13}^2+c_{13}^2c_{12}^2),\\
B=4c_{12}s_{12}c_{23}s_{23}s_{13}.
\eea

The novel correlations obtained in Eq.\ref{cd} are exact and can be further simplified if terms $\mathcal{O}(s^4_{13})$ are dropped. Interestingly, both the relations are independent of $\theta_g$ and coincide with the prediction $\cos\delta=0$ for CP extended $\mu\tau$ ($\rm CP^{\mu\tau}$)\cite{H1,H2} in the limit $\theta_{23}\to\pi/4$.
\begin{figure}[H]
\includegraphics[scale=.55]{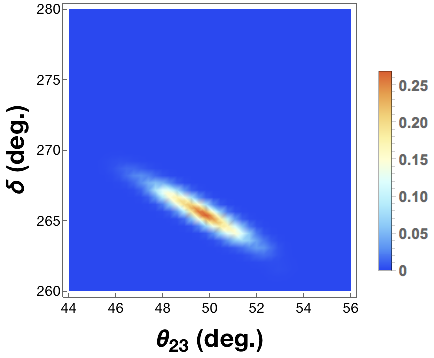} \includegraphics[scale=.54]{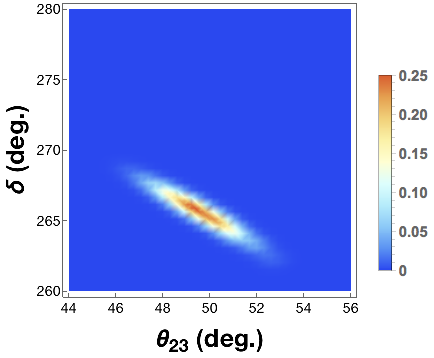}\\
\includegraphics[scale=.55]{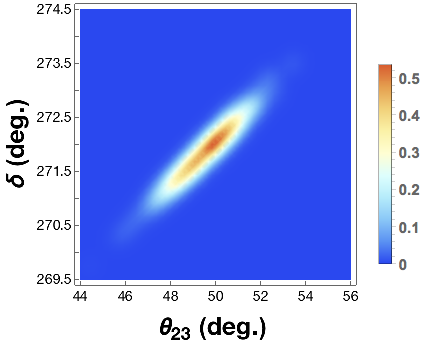} \includegraphics[scale=.54]{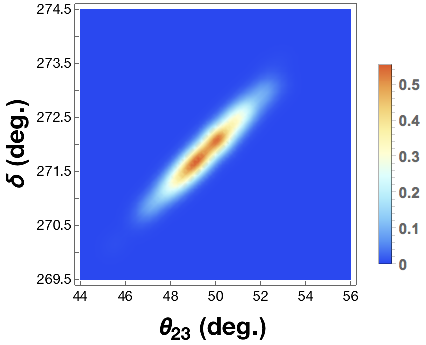}
\caption{ For the real invariance $ (\mathcal{G}_{1}^{g\mu\tau})^TM_\nu \mathcal{G}_{1}^{g\mu\tau}=M_\nu$ (top): Probability distribution of $\delta$ with $\theta_{23}$. For the real invariance $ (\mathcal{G}_{2}^{g\mu\tau})^TM_\nu \mathcal{G}_{2}^{g\mu\tau}=M_\nu$ (bottom):  Probability distribution of $\delta$ with $\theta_{23}$. Here we have used Gaussian distribution for each of the mixing angles.}\label{fig5}
\end{figure}  

In Fig.\ref{fig5}, we show the predictions of CP extended $\mu\tau$ mixing. The figures in the top panel  are for the real invariance for $\mathcal{G}_1^{g\mu\tau}$ (for each case there are two solutions due the `$\pm$' sign in Eq.\ref{cd}) and those which are in the bottom panel are for the real invariance due to $\mathcal{G}_2^{g\mu\tau}$. Notice that unlike the $\rm CP^{\mu\tau}$ (CP extended $\mu\tau$ interchange\cite{H3}) which predicts co-bimaximal mixing ($\delta=\pm 3\pi/2$ and $\theta_{23}=\pi/4$) here nonmaximal atmospheric mixing is allowed, however, the most probable values of $\delta$ are clustered around their near maximal values. Thus significant deviation from maximality of $\delta$ would rule out both the scenarios (present data on $\delta$ for NMO is at $\sim 1.37\sigma$ tension with the predictions obtained in the CP extended $\mu\tau$ mixing). Nevertheless, we should stress that since both the real invariances lead to almost same predictions on $\delta$ (a preference for near maximality), it is difficult to disentangle these two cases experimentally, unless we look for precision measurements of $\delta$ in near future or we do explicit mass models which can be testable through other parameters, e.g, through predictions on masses or neutrino-less double beta decay parameters etc. Before proceeding further into the discussion of the CP extended $\mu\tau$ mixing in Type-I seesaw framework with the motivation to explore possible implications on baryogenesis via  leptogenesis, let's point out an interesting aspect regarding the class of the CP extended $\mu\tau$ mixing within the proposed  $\mu\tau$ mixing.   We have seen in Sec.\ref{s2}, introduction of $\mu\tau$ mixing instead of the $\mu\tau$ interchange symmetry, adds up two more degrees of freedom. To be precise, for both the cases, whilst simultaneous nonmaximal values for $\delta$ and $\theta_{23}$ are allowed,  only for the $\mu\tau$ mixing scenario, we can test nonmaximality in one parameter for a maximal value of the other. Therefore there could be two prominent options to look for i) $\delta$ could be maximal but $\theta_{23}$ is not ii) $\theta_{23}$ could be maximal but $\delta$ is not. From Fig.\ref{fig5}, it is evident that the CP extended $\mu\tau$ mixing belongs (approximately) to the case `i'. Explicit flavour models for the case `ii' would also be interesting in future since we don't have any precise statements on the value of $\delta$ and $\theta_{23}$ at this moment.\\
\subsection{CP symmetry in Type-I seesaw}
We may now proceed to the discussion of this extended CP in Type-I seesaw mechanism.  It is well known that to obtain light neutrino masses one has to introduce singlet right handed (RH) fields  as a minimal extension to the  Standard Model (SM). Thus with the introduction of the singlet fields $N_{Ri}$, in the diagonal basis of the RH neutrinos, the   Lagrangian for the Type-I seesaw reads
\bea
-\mathcal{L}^\nu_{\rm mass}= \bar{N}_{iR} (m_D)_{i\alpha}\nu_{L\alpha}+\frac{1}{2}\bar{N}_{iR}(M_R)_i \delta _{ij}N_{jR}^C + {\rm h.c.}\label{selag}
\eea
with $\alpha=e,\mu,\tau$ and $i=1,2,3$.
The first term in Eq.\ref{selag} is a Dirac type and the second term is a Majorana type mass term which together lead to the effective $3\times3$ light neutrino Majorana mass matrix $M_\nu$ as
\bea
M_\nu=-m_D^TM_R^{-1}m_D. \label{swmnu}
\eea
In the diagonal basis of the charged lepton as well as the heavy RH neutrinos, a CP invariant light neutrino mass matrix 
\bea
\mathcal{G}^TM_\nu\mathcal{G}=M_\nu^*
\eea
could be obtained with the following transformation\footnote{We choose the minus sign in the RHS of Eq.\ref{minus}, just to present entries of the first column of \ref{m_D} as real elements. Otherwise, given the form $\mathcal{G}_3^{g\mu\tau}$ in Eq.\ref{mixG}, those elements would be pure imaginary quantities. However, this is just a matter of convention. A plus sign is also equally acceptable.} on $m_D$.
\bea
m_D \mathcal{G}=-m_D^* \label{minus}
\eea
We refer to Refs.\cite{H3,CP4,Chen:2016ptr}, to realize how in the diagonal basis of  charged lepton and heavy neutrinos, CP is applied to the Type-I seesaw Lagrangian. Now in our case, to have a real invariance for $\mathcal{G}_1^{g\mu\tau}$ as well as  $\mathcal{G}_2^{g\mu\tau}$ one needs the following transformations.
\bea
m_D \mathcal{G}_i^{g\mu\tau}=-m_D^*, m_D \mathcal{G}_3^{g\mu\tau}=-m_D^*.\hspace{1cm} (i=1,2)\label{cptrans}
\eea
For both the cases, the most general form of $m_D$ that satisfies the second constraint of Eq.\ref{cptrans} can be parametrized as 
\begin{equation}
m_D=\begin{pmatrix}
a & b_1+ib_2 & -b_1\tan\theta_g+ib_2\cot\theta_g\\
e & c_1+ic_2 & -c_1\tan\theta_g+ic_2\cot\theta_g\\
f & d_1+id_2 & -d_1\tan\theta_g+id_2\cot\theta_g
\end{pmatrix},\label{m_D}
\end{equation}
where all the parameters are real and a priori unknown.
There will be other constraints (the parameters $b_1,c_1,d_1$ could be expressed in terms $a,e,f$ and $\theta_g$) on the mass matrix $m_D$ of Eq.\ref{m_D} due to the first transformation in Eq.\ref{cptrans}. However, those transformations are not important in this paper, since we  present the discussion of leptogenesis with few benchmark values (particularly for the decay parameters $K_{i\alpha}$ and  the CP asymmetry parameters $\epsilon_{i\alpha}$ as given in the next section) which are always compatible with those transformations\footnote{Even after considering the constraints from $\mathcal{G}_{1,2}^{g\mu\tau}$, the number of effective parameters in $m_D$ are more than the number of experimental constraints.}. In any case, those constraint equations could easily be derived as shown in Ref.\cite{CP4}.    But what matters here, the overall structure of $m_D$ shown in Eq.\ref{m_D}. Having set up all the necessary prerequisites, we are now ready to explore the baryogenesis via leptogenesis in the CP extended $\mu\tau$ mixing framework.
\section{Baryogenesis via leptogenesis in the  CP extended framework}\label{s4}
Baryogenesis via leptogenesis is a process where CP violating and out of equilibrium decays of the heavy RH neutrinos produce lepton asymmetry which is thereafter converted to baryon asymmetry by non-perturbative sphalerons\cite{sph}. For a simplified discussion we opt for a two RH neutrino scenario where the decays and interactions due to $N_1$ and $N_2$ would matter to the process of leptogenesis. However, the qualitative results drawn in a two RH neutrino scenario would also be relevant for a three RH neutrino case. 
When the masses of the two RH neutrinos are in the regime $M_i >10^{12}$ GeV where all the charged lepton flavours are out of equilibrium\cite{lep1,lep2,lep3},  quantum states $\ket{\ell_i}$ produced by the decay of RH neutrinos can be written as a coherent superposition of the flavour states $\ket{\ell_{\alpha}}$ as
\bea
\ket{\ell_i}&=&\mathcal{A}_{i\alpha} \ket{\ell_\alpha}, \hspace{1cm} \\
\ket{\bar{\ell}_i}&=&\bar{\mathcal{A}}_{i\alpha} \ket{\bar{\ell}_\alpha}. \hspace{1cm} (i=1,2, \alpha=e,\mu,\tau)
\eea
 The amplitudes at the tree level are given by
   \bea
 \mathcal{A}_{i\alpha}^0 =\frac{m_{D_{i\alpha}}}{\sqrt{(m_Dm_D^\dagger)_{ii}}}\hspace{1cm}{\rm and}\hspace{1cm}\bar{\mathcal{A}}_{i\alpha}^0 =\frac{m^*_{D_{i\alpha}}}{\sqrt{(m_Dm_D^\dagger)_{ii}}}.
 \eea  
    Since there is no interaction to break the coherence of the quantum states before it inversely decays to $N_i$, the asymmetry will be produced along the direction of $\ket{\ell_i}$(or  $\ket{\bar{\ell}_i}$) in the lepton flavour space. In that case, the set of classical kinetic equations relevant for leptogenesis could be written as\cite{L3}
\bea
\frac{dN_{N_i}}{dz}&=&-D_i(N_{N_i}-N_{N_i}^{\rm eq}), ~{\rm with}~i=1,2 \label{be1}\\
\frac{dN_{B-L}}{dz}&=&-\sum_{i=1}^2\varepsilon_i D_i(N_{N_i}-N_{N_i}^{\rm eq})-\sum_{i=1}^2W_iN_{B-L},\label{be2}
\eea
 with $z=M_1/T$. Eq.\ref{be1} tracks the dynamics of the RH neutrinos (production+decay) while Eq.\ref{be2} tracks the lepton asymmetry which survives in the interplay of the production (first term) and washout (second term),  as a function of $z$.  $N_{N_i}$'s and $N_{B-L}$ are the abundances computed per number of $N_i$'s in ultra-relativistic thermal equilibrium. Defining $x_{ij}=M_j^2/M_i^2$ and $z_i=z\sqrt{x_{1i}}$, , the decay terms can be written as
\bea
D_i=\frac{\Gamma_{D,i}}{H z}=K_ix_{1i}z   \langle 1/\gamma_i\rangle,
\eea
where total decay rates $\Gamma_{D,i}$ are given  by $\Gamma_{D,i}=\bar{\Gamma}_i+\Gamma_i=\Gamma_{D,i}(T=0) \langle 1/\gamma_i\rangle$ with $ \langle 1/\gamma_i\rangle$ as the thermally averaged dilation factor and   can be expresses as the ratio of two modified Bessel functions as
\bea
 \langle 1/\gamma_i\rangle=\frac{\mathcal{K}_1(z_i)}{\mathcal{K}_2(z_i)}.
\eea
 The decay parameter $K_i$ is given by
  \bea 
 K_i\equiv\Gamma_{D,i}(T=0)/H(T=M_i).
 \eea
 The equilibrium abundance of $N_i$ is  given by $N_{N_i}^{\rm eq}=\frac{1}{2} z_i^2\mathcal{K}_2(z_i)$ and the  CP asymmetry $\varepsilon_i=\sum_{\alpha}\varepsilon_{i\alpha}$ is given by
 \bea
 \varepsilon_{i}=\sum_{\alpha}\frac{\Gamma_{i\alpha}-\bar{\Gamma}_{i\alpha}}{\Gamma_i+\bar{\Gamma}_i}\label{epsi}
 \eea 
 with the flavoured CP asymmetry $\varepsilon_{i\alpha}$ as\cite{Adhikary:2014qba}
\bea
\varepsilon_{i\alpha}
&=&\frac{1}{4\pi v^2 h_{ii}}\sum_{j\ne i} {\rm Im}\{h_{ij}
({m_D})_{i\alpha} (m_D^*)_{j\alpha }\}
\left[f(x_{ij})+\frac{\sqrt{x_{ij}}(1-x_{ij})}
{(1-x_{ij})^2+{{h}_{jj}^2}{(16 \pi^2 v^4)}^{-1}}\right]\nonumber\\
&+&\frac{1}{4\pi v^2 {h}_{ii}}\sum_{j\ne i}\frac{(1-x_{ij})
{\rm Im}\{{h}_{ji}({m_D})_{i\alpha} (m_D^*)_{j\alpha}\}}
{(1-x_{ij})^2+{{h}_{jj}^2}{(16 \pi^2 v^4)}^{-1}},\label{ncp}
\eea 
where $h_{ij}\equiv( m_D m_D^\dag)_{ij}$. The final $B-L$ asymmetry could be written as
\bea
N_{B-L}^f=N_{B-L}^{\rm in} e^{-\sum_{i}\int dz^\prime W_i(z^\prime)}+N_{B-L}^{\rm lepto},\label{flepto}
\eea
where $N_{B-L}^{\rm in}$ could be a possible pre-existing asymmetry at an initial temperature $T_{\rm in}$ and $N_{B-L}^{\rm lepto}$ is the contribution from pure leptogenesis.  In this work we assume any pre-existing asymmetry (so called strong thermal condition\cite{G8}) is strongly washed out by the heavy RH neutrinos. Therefore, we are in a strong washout scenario. Thus in the washout term in Eq.\ref{be2},  the $\Delta L=1$ scattering term $W^{\Delta L=1}_i$  can be safely neglected\cite{lep2}. However, a particular washout regime is a matter of choice in our discussion. One can neglect the pre-existing asymmetry assuming there is no source of asymmetry production prior to the   leptogenesis phase and explore a weak washout regime as well. In that case inclusion of scattering would only affect the asymmetry production efficiency\cite{lep2}  but the qualitative conclusion drawn in a strong washout regime would remain the same.  For our purpose, we shall also neglect the   non-resonant part of the $\Delta L=2$ term $W^{\Delta L=2}_i$ which is relevant only at higher temperature. Now the relevant washout term $W_i\simeq W_i^{\rm ID}$ can be written as (after properly subtracting  the real intermediate state contribution of $\Delta L=2$ process)
\bea
W_i^{\rm ID}=\frac{1}{4}K_i\sqrt{x_{1i}}\mathcal{K}_1(z_i)z_i^3.\label{inv_decay}
\eea
 The final baryon to photon ratio is  given by
\bea
\eta_B=a_{\rm sph}\frac{N_{B-L}^{\rm lepto}}{N_\gamma^{\rm rec}}\simeq 0.96\times 10^{-2}N_{B-L}^{\rm lepto},
\eea
where $N^{\rm rec}_\gamma$ is the photon density at the recombination and the sphaleron conversion coefficient $a_{\rm sph}\sim 1/3$. In a given model, this $\eta_{B}$ has to be compared with measured value\cite{planck}
\bea
\eta_{B}^{\rm CMB}=(6.3\pm 0.3)\times 10^{-10}. \label{etaB}
\eea
In the mass regime $10^{9}$ GeV $<M_i<10^{12}$ GeV, interactions due to tau charged lepton flavour are fast enough to break the coherent evolution of the quantum states $\ket{\ell_i}$ before it inversely decays to $N_i$.  The $\ket{\ell_i}$ is then projected into a two flavour basis characterised by the eigenstates along  the directions of $\tau$ and $\tau_{i}^\perp=e+\mu$. In the three flavour regime, i.e. $M_i<10^{9}$ GeV,  the muon charge lepton flavour  comes into equilibrium. It breaks the coherent evolution of the states which is along $\tau_{i}^\perp$ and  one resolves all the flavours ($e,\mu,\tau$) individually (for both the flavour regimes, we are assuming strong decoherence so that the density matrix\footnote{An elaborate computation of leptogenesis in density matrix formalism is given in Ref.\cite{Moffat:2018wke}} is flavour diagonal\cite{lep1,bari}). Thus for each flavour regime, one has to track the lepton asymmetry in the relevant flavours. For example, if we are in the two flavour regime, the lepton asymmetry has to be tracked in $\tau$ and $\tau_i^\perp$ flavours. The Boltzmann equations for a generic flavour `$\alpha$' could be written as
\bea
\frac{dN_{N_i}}{dz}&=&-D_i(N_{N_i}-N_{N_i}^{\rm eq}), ~{\rm with}~i=1,2 \label{bef1}\\
\frac{dN_{{\Delta}_{\alpha}}}{dz}&=&-\sum_{i=1}^2\varepsilon_{i\alpha} D_i(N_{N_i}-N_{N_i}^{\rm eq})-\sum_{i=1}^2P_{i\alpha}^0W_i^{\rm ID}N_{{\Delta}_{\alpha}}.\label{bef2}
\eea
Here $N_{\Delta_{\alpha}}$ is the asymmetry in the flavour $\alpha$ analytic solution for which can be obtained as
\bea
N_{\Delta_{\alpha}}=-\sum_{i}^2\varepsilon_{i\alpha}\kappa_{i\alpha}\label{fep}
\eea
with the efficiency factor
\bea
\kappa_{i\alpha} (z)=-\int_{z_{\rm in}}^\infty \frac{dN_{N_i}}{dz^\prime}e^{-\sum_{i}\int_{z^\prime}^zP_{i\alpha}^0 W_{i}^{\rm ID}(z^{\prime\prime})dz^{\prime\prime}}dz^\prime. \label{effi}
 \eea
For numerical integration purposes one can set very small value of `$z_{\rm in}(\sim 0)$' and a very large value for `$z(\sim 10^3)$'. The final baryon to photon ratio is then given by 
\bea
\eta_{B}=0.96\times10^{-2}\sum_{\alpha}N_{\Delta_{\alpha}}.
\eea
The quantity $P_{i\alpha}^0$ is the tree level probability of a quantum state produced by the $i$th heavy neutrino being in the flavour $\alpha$ and has an expression 
\bea
P_{i\alpha}^0\equiv K_{i\alpha}/K_i,
\eea
where $K_{i\alpha}$ is the flavoured decay parameter defined as 
\bea
 K_{i\alpha}=\frac{\Gamma_{i\alpha}+\bar{\Gamma}_{i\alpha}}{H(T=M_i)}\equiv\frac{|m_{D_{i\alpha}}|^2}{M_i m^*}\label{fdecay}
 \eea
 with $m^*\simeq 10^{-3}$ being the equilibrium neutrino mass. Let us mention another important parameter $\Delta P_{i\alpha}=P_{i\alpha}-\bar{P}_{i\alpha}$ strongly relevant to our discussion. The tree+loop level projectors are given by 
 $P_{i\alpha}=|\mathcal{A}_{i\alpha}|^2=P_{i\alpha}^0+\frac{\Delta P_{i\alpha}}{2}$, $\bar{P}_{i\alpha}=|\bar{\mathcal{A}}_{i\alpha}|^2=P_{i\alpha}^0-\frac{\Delta P_{i\alpha}}{2}$. The quantity $\Delta P_{i\alpha}$, the difference between the tree+loop level projectors, is  nonzero  since, in general $|\mathcal{A}_{i\alpha}|\neq |\bar{\mathcal{A}}_{i\alpha}|$\cite{L2p}. Now the flavoured CP asymmetry parameter $\varepsilon_{i\alpha}$ of Eq.\ref{epsi} can be simplified as
 \bea
 \varepsilon_{i\alpha}=P_{i\alpha}^0\varepsilon_i+\Delta P_{i\alpha}/2.
 \eea
Though  the quantity $\Delta P_{i\alpha}$ is not significant in the washout terms,  for  the CP asymmetry parameter it is remarkably relevant. In fact we show in the model under consideration, the entire source of CP violation in a particular flavour `$\alpha$' arises due to the $\Delta P_{i\alpha}$ term.\\

As mentioned in the introduction, here we discuss only the three flavour regime ($M_i<10^{9}$ GeV) of the leptogenesis to show the dramatic difference between the conclusion drawn in case of a CP extended $\mu\tau$ interchange\cite{H3,H5,Chen:2016ptr} and the proposed CP extended $\mu\tau$ mixing. Let's clarify explicitly why we do that. \\

{\it One flavour regime:} First of all, for the one flavour regime ($M_i>10^{12}$ GeV), the second term in Eq.\ref{ncp} is vanishing when summed over `$\alpha$', i.e, ${\rm Im}\{{h}_{ji}({m_D})_{i\alpha} (m_D^*)_{j\alpha}\}={\rm Im}[|h_{ji}|^2]=0$. The first term is proportional to ${\rm Im}\{{h}_{ij}^2\}$. Using Eq.\ref{m_D}, the `$h=m_Dm_D^\dagger$' can be shown to be a real matrix. Thus the flavour summed CP asymmetry $\varepsilon_i=\sum_{\alpha}\varepsilon_{i\alpha}$ vanishes for any `$i$'.  Therefore successful baryogenesis is not possible in the unflavoured regime. This result has also been  obtained in CP extended $\mu\tau$ interchange symmetry\cite{H3,H5,Chen:2016ptr}. One can also show $\varepsilon_{ie}=0$, since the first column of the $m_D$ matrix in Eq.\ref{m_D} is real. Thus similar to CP extended $\mu\tau$ interchange, $\varepsilon_{i\mu}\equiv\Delta P_{i\mu}/2=-\varepsilon_{i\tau}$. Therefore, in the one flavour regime,  the results obtained  for leptogenesis in  CP extended $\mu\tau$ mixing, are similar to  CP extended $\mu\tau$ interchange. \\

{\it Two flavour regime:} As already mentioned, we are probing a strong washout scenario assuming any pre-existing asymmetry produced prior to leptogenesis phase has been strongly washed out. This is only possible in the three flavour regime. For the two flavour regime  ($10^{9}$ GeV $<M_i<10^{12}$ GeV), note that this is not possible since, though in the direction of $\tau$ flavour the asymmetry could be washed out assuming $K_{i\tau}\gg1$, however, a component of the asymmetry would always survive in the direction orthogonal to the $\tau_\perp$\cite{Engelhard:2006yg,bari}, irrespective of the value of $K_{\tau_\perp}$. Thus a pure leptogenesis scenario breaks down. In any case,  as mentioned earlier,  along with the strong washout scenario, one can also probe the weak washout regime relaxing the strong-thermal condition (pre-existing asymmetry), which has been done so far in the literature in the context of $\rm CP^{\mu\tau}$). However, for the latter case, apart form showing a successful baryogenesis, we hardly  expect any prediction on low energy neutrino parameters, since  the number of parameters in $m_D$ is still larger than the number of experimental constraints. Thus in the two flavour regime, from leptogenesis perspective, there will be no significant difference between a CP extended $\mu\tau$ interchange and a CP extended $\mu\tau$ mixing.  But certainly differences will be there if one assumes texture zeros on top of the CP extended $\mu\tau$ mixing\cite{Branco:2007nb,mu10,Barreiros:2018bju} or imposes the symmetry in a minimal seesaw framework\cite{H5,Barreiros:2018bju}. Since in that case there would be less number of parameters and one might expect predictions from the baryogenesis  constraint on the physical parameters such as $\theta_{23}$  which is nonmaximal in general in the $\mu\tau$ mixing scheme. However, an elaborate description in this context, is beyond the scope of this paper.\\

\begin{figure}[H]
\begin{center}
\includegraphics[scale=.45]{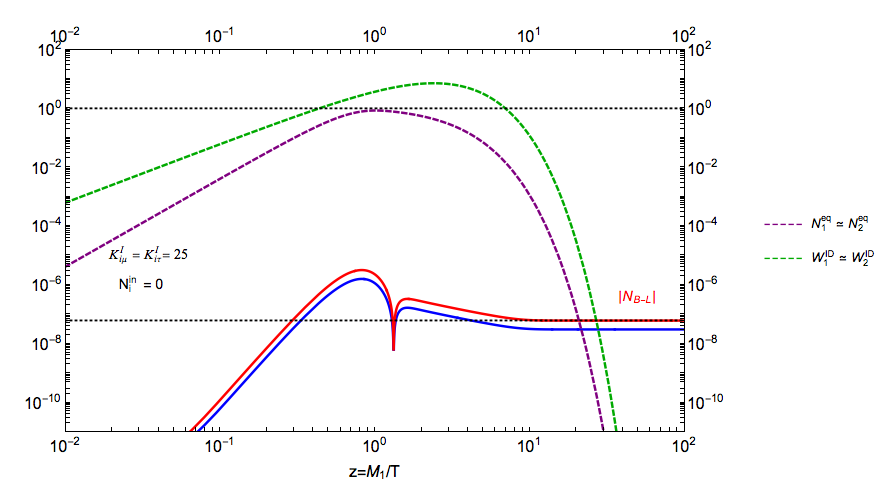} 
\caption{ Variation of $N_{B-L}$ with $z$ assuming $4\delta_xK_{i\tau}^{I} \cos 2\xi=0.1$ and $|\varepsilon_{i\tau}|\simeq 1.5\times 10^{-3}$. The blue line is the contribution from a single RH neutrino. The red line which matches the observed range, represents contributions from both the RH neutrinos.}\label{fig6}
\end{center}
\end{figure}

{\it Three flavour regime:} Now coming back to the discussion of leptogenesis in the three flavour regime, first of all one has to go beyond the hierarchical scenario, since in the hierarchical limit the CP asymmetry parameter, say, $\varepsilon_{1\alpha}$ is proportional to $M_1$ and if $M_1<10^{9}$ GeV, one can not generate the observed baryon asymmetry\cite{L2pp}. However if the mass differences of the RH neutrinos are close enough, one expects a significant enhancement in the loop functions, particularly self energy  contribution to the CP asymmetry parameter increases and therefore even if $M_i<10^{9}$ GeV, required baryon asymmetry  could be generated due to this enhancement in the CP asymmetry parameter\cite{reso}.
However, in that scenario one has to consider the asymmetry generated by all the heavy neutrinos, since in the standard hierarchical scenario, contribution from the heavier RH neutrinos are washed out by the lighter RH neutrinos. In the limit of quasi-degeneracy (QD) in the RH neutrino spectrum, the contribution from the heavier neutrinos can not be washed out. For an explicit analytical explanation of leptogenesis due to QD mass spectrum, we refer to\cite{Samanta:2018hqm}. For the CP extended $\mu\tau$ interchange symmetry as well as mixing, the $N_{B-L}$ asymmetry could be written as 
\bea
N_{B-L}^f=\sum_\alpha N_{\Delta_{\alpha}}=-\sum_{i}^2\left(\varepsilon_{i\tau}\kappa_{i\tau}+\varepsilon_{i\mu}\kappa_{i\mu}\right)=-\sum_{i}^2\varepsilon_{i\tau}\left(\kappa_{i\tau}-\kappa_{i\mu}\right)=-\sum_{i}^2\varepsilon_{i\tau}\kappa_{i}^{\rm eff},\label{bar1}
\eea
where we use the fact that $\varepsilon_{ie}=0$ and $\varepsilon_{i\mu}=-\varepsilon_{i\tau}$
and at $z\rightarrow\infty$, the efficiency factor $\kappa_{i\alpha}$ has the expression 
\bea
\kappa_{i\alpha} =-\int_{0}^\infty \frac{dN_{N_i}}{dz^\prime}e^{-\sum_{i}\int_{z^\prime}^{\infty}(K_{i\alpha}/K_i) W_{i}^{\rm ID}(z^{\prime\prime})dz^{\prime\prime}}dz^\prime \hspace{1cm}\alpha=(\tau,\mu). \label{effi2}
 \eea
Now notice that, for CP extended $\mu\tau$ interchange ($\theta_g\rightarrow \pi/4$), using Eq.\ref{m_D} and Eq.\ref{fdecay} the decay parameters can be obtained as
\bea
K_{1\mu}^{I}=\frac{b_1^2+b_2^2}{M_1 m^*}=K_{1\tau}^{I} ,\hspace{.5cm} K_{2\mu}^{I}=\frac{c_1^2+c_2^2}{M_2 m^*}=K_{2\tau}^{I},\label{fdp}
\eea
where `$I$' stands for `Interchange'. Thus from Eq.\ref{effi2} one concludes $\kappa_{i\mu}=\kappa_{i\tau}$ and hence, from  Eq.\ref{bar1}, $N_{B-L}^f=0$. Therefore, even if we are in the resonance regime of leptogenesis, baryon asymmetry vanishes due to the exact cancellation of the efficiency factors. But for the CP extended $\mu\tau$ mixing this is not the case. This is since, though the decay parameters $K_{i\mu}$  have the same expression as shown in Eq.\ref{fdp}, since $\theta_g\neq 0$ in general, $K_{i\tau}^M$ can be obtained as 
\bea
K_{i\tau}^{M}=\frac{[{\rm Re}(m_D)_{i\mu}]^2\tan^2\theta_{g}+[{\rm Im}(m_D)_{i\mu}]^2\cot^2\theta_{g}}{M_im^*}
\eea
which reduces to $K_{i\tau}^I$ of Eq.\ref{fdp} in the limit $\theta_g\rightarrow \pi/4$. Here `$M$' stands for `Mixing'. In fact, given the distribution of $\delta$ in Fig.\ref{fig5}, we can approximate $\sin\delta\sim 1$ and therefore using Eq.\ref{g3c}, we can recast $K_{i\tau}^M$ as 
\bea
K_{i\tau}^{M}=\frac{[{\rm Re}(m_D)_{i\mu}]^2\tan^2\theta_{23}+[{\rm Im}(m_D)_{i\mu}]^2\cot^2\theta_{23}}{M_im^*}. \label{decth}
\eea
Thus since $K_{i\tau}^{M}\neq K_{i\mu}^{M}$, from Eq.\ref{bar1} $N_{B-L}$ is nonvanishing. It is now clear that to obtain a nonzero baryon asymmetry, in this CP extended $\mu\tau$ mixing framework, one always needs deviation of $\theta_{23}$ from maximality. Now parametrizing $\theta_{23}$ as $\theta_{23}=(\pi/4+\delta_x)$, where the parameter  $\delta_x$ accounts for the nonmaximality of $\theta_{23}$, Eq.\ref{decth} could further be simplified as
\bea
K_{i\tau}^{M}=K_{i\tau}^{I}(1+4\delta_x \cos 2\xi_i),\hspace{1cm}\xi_i=\tan^{-1}\frac{{\rm Im}[(m_D)_{i\mu}]}{{\rm Re}[(m_D)_{i\mu}]}.
\eea
Thus barring a very special solution $\cos 2\xi_i=0$, one always obtains a nonvanishing baryon asymmetry for a nonmaximal value of $\theta_{23}$. Now the  $\kappa_{i}^{\rm eff}$ of Eq.\ref{bar1} can be obtained as 
\bea
\kappa_{i}^{\rm eff}=4\delta_xK_{i\tau}^{I} \cos 2\xi_i\int_{0}^\infty \frac{dN_{N_i}}{dz^\prime}e^{-\sum_{i}\int_{z^\prime}^{\infty}K_{i\tau}^I K_i^{-1}W_{i}^{\rm ID}(z^{\prime\prime})dz^{\prime\prime}}\sum_{i}\int_{z^\prime}^{\infty} K_i^{-1}W_{i}^{\rm ID}(z^{\prime\prime})dz^{\prime\prime}dz^\prime.\label{efficp}
\eea
For convenience we may choose $\cos2\xi_i\equiv \cos2\xi=1$ \footnote{This is always not the case. The parameter $\xi$ is model dependent. However, as we have already pointed out earlier, the models with CP symmetries, one can not constrain the mass matrix element only by oscillation data, unless some special conditions are assumed\cite{mu10}. Thus $\cos 2 \xi=1$ would be a probable solution.}  and flavoured decay parameters for both the RH neutrinos are the same. Then we may further parametrise the $\kappa_{i}^{\rm eff}$ as 
\bea
\kappa_{i}^{\rm eff}\leq\overline{m}_{\rm max}\frac{4\delta_x}{m^*}\int_{0}^\infty \frac{dN_{N_i}}{dz^\prime}e^{-\sum_{i}\int_{z^\prime}^{\infty}K_{i\tau}^I K_i^{-1}W_{i}^{\rm ID}(z^{\prime\prime})dz^{\prime\prime}}\sum_{i}\int_{z^\prime}^{\infty}K_i^{-1} W_{i}^{\rm ID}(z^{\prime\prime})dz^{\prime\prime}dz^\prime, \label{efficp2}
\eea
where $\overline{m}$ is the overall mass scale of the light neutrinos. The effective efficiency factor in Eq.\ref{efficp2} has very interesting features. First of all, from the perspective of $\kappa_{1}^{\rm eff}$ (effective production efficiency due of $N_1$), it encounters a two step suppression. The first one is due to the $N_2$-washout, since unlike the hierarchical scenario, in the QD limit $N_2$ washout significantly reduces the asymmetry produced by $N_1$\cite{Samanta:2018hqm} and the second one is due the $\delta_x$ which appears as a pre-factor in Eq.\ref{efficp2}. In case of $\kappa_{2}^{\rm eff}$, firstly it increases since the $N_1$ interactions cannot fully washout the  asymmetry produced by $N_2$ (the production from $N_2$ is still on), however again similar to $\kappa_{1}^{\rm eff}$, it faces a suppression by $\delta_x$. However, due to the small mass splitting between the RH neutrino masses, at the end, one obtains comparable production efficiencies ($\kappa_{1}^{\rm eff}\simeq \kappa_{2}^{\rm eff}$). Let us now have a numerical estimate of the final baryon asymmetry. From Eq.\ref{etaB}, it is clear that $N_{B-L}=\sum_{\alpha}N_{\Delta_\alpha}\simeq 6.3 \times 10^{-8}$. Now if we choose a very small value of the pre-factor in Eq.\ref{efficp}, say, $4\delta_xK_{i\tau}^{I} \cos 2\xi_i \simeq 0.1$ (this can be done, e.g., either by choosing a very small value of $\cos 2\xi_i$ or very small value of $\delta_x$), a numerical integration of Eq.\ref{efficp} gives $\kappa_{i}^{\rm eff}\simeq 2\times 10^{-5}$, where we have assumed $K_{i\mu}=K_{i\tau}=25$. Therefore, to be consistent with the observed value of $N_{B-L}$,  one needs $|\varepsilon_{i\tau}|\simeq 1.5\times 10^{-3}$. This has also been reproduced in Fig.\ref{fig6} by solving the Boltzmann equations, assuming both the RH neutrinos contribute equally (this is justified when one chooses very small mass splitting between RH neutrinos, which is needed in this scheme to obtain resonance in the three flavour regime). One can also comment on the mass scale of the RH neutrinos. For example, let say the resonant enhancement of the CP asymmetry happens when the mass difference $\Delta=\sqrt{x_{12}}-1\simeq 10^{-7}$. Then assuming the elements of $m_D$ in Eq.\ref{ncp} as $m_D\sim \sqrt{M\overline{m}}$ one obtains the mass scale of the RH neutrinos $M\sim 10^7$ GeV. One can explore another interesting situation, assuming $\cos 2\xi =1$, $\overline{m}_{\rm max}\simeq \sqrt{|\Delta m_{23}^2|}$ and the current best-fit of $\theta_{23}\simeq 49.6^0~(\delta_x\sim 4.6^0)$.  This would correspond to the value of the pre-factor in Eq.\ref{efficp2} as $\overline{m}_{\rm max}\frac{4\delta_x}{m^*}\simeq 8$.  In that case the correct value of $\eta_B$ could be generated with $|\varepsilon_{i\tau}|\simeq 2.6\times 10^{-5}$ and consequently the mass scale of the RH neutrinos could be lowered to $\sim$100 TeV. 
Having established the possibility of resonant leptogenesis in the CP extended $\mu\tau$ mixing scheme, the main purpose of the leptogenesis study in this paper is served. However, still one would like to consider some other interesting possibilities such as flavour coupling\cite{lep4,lep5,lep6}. So far we have discussed the leptogenesis scenario without flavour coupling matrix\cite{lep3} in the Boltzmann equations. With flavour couplings the Boltzmann equation of Eq.\ref{bef1} and Eq.\ref{bef2} will be modified as 
\bea
\frac{dN_{N_i}}{dz}&=&-D_i(N_{N_i}-N_{N_i}^{\rm eq}), ~{\rm with}~i=1,2 \label{bef3}\\
\frac{dN_{{\Delta}_{\alpha}}}{dz}&=&-\sum_{i=1}^2\varepsilon_{i\alpha} D_i(N_{N_i}-N_{N_i}^{\rm eq})-\sum_{i=1}^2P_{i\alpha}^0W_i^{\rm ID}\sum_{\beta = e, \mu, \tau}C_{\alpha\beta}N_{{\Delta}_{\beta}},\label{bef4}
\eea
where the flavour coupling matrix $C_{\alpha\beta}$ is given by
\bea
C_{\alpha\beta}=\begin{pmatrix}
188/179&32/179&32/179\\
49/358&500/537&142/537\\
49/358&142/537&500/537
\end{pmatrix}
\eea
which properly accounts for the asymmetry in lepton doublets as well as Higgs asymmetry.
One might  wonder  whether the flavour coupling effect can save the situation for the CP extended $\mu\tau$ interchange symmetry, i.e., whether the entries of the flavour coupling can create a mismatch between the decay parameters so that one obtains a nonzero $\kappa_{i}^{\rm eff}$. Starting from the simplest case, i.e., assuming the diagonal structure of the $C$ matrix (which is usually used in the leptogenesis computations for neutrino mass models), if one writes the Boltzmann equations, the scenario does not change. This is since the $C_{\mu\mu}$ and $C_{\tau\tau}$ elements are the same and therefore they are unable to create any mismatch between the decay parameters. Thus similar to the previous scenario ($C=\mathbb{I}$)  there will be no net lepton asymmetry. Interestingly, even if one assumes the nondiagonal structure of the $C$ matrix, one cannot generate nonzero lepton asymmetry. Since, $P^0_{i\mu}=P^0_{i\tau}$, the muon and tau flavour will couple each other with equal strength ($142 P^0_{i\mu}/537$). Therefore there will be no net asymmetry mismatch since we already know the production is equal an opposite ($\varepsilon_{i\mu}=-\varepsilon_{i\tau}$). Mathematically, this can be understood in the following way. We can go to a basis where the Boltzmann equation in Eq.\ref{bef4} is diagonal in a generic flavour space, say $\alpha^\prime$. We can do this by the means of a unitary transformation as
\bea
\frac{dN_{{\Delta}_{\alpha^\prime}}}{dz}&=&-\sum_{i=1}^2\varepsilon_{i\alpha^\prime} D_i(N_{N_i}-N_{N_i}^{\rm eq})-\sum_{i=1}^2W_i^{\rm ID}V^{-1}P_{i\alpha}^0C_{\alpha\beta}VN_{{\Delta}_{\beta^\prime}},\label{bef6}
\eea
where 
\bea
N_{{\Delta}_{\beta^\prime}}=V^{-1}N_{{\Delta}_{\beta}},\hspace{.5cm}\varepsilon_{i\alpha^\prime}=V^{-1}\varepsilon_{i\alpha},\hspace{.5cm} V^{-1}P_{i\alpha}^0C_{\alpha\beta}V =P_{i\beta^\prime}^0\delta_{\alpha^\prime\beta^\prime}.
\eea
  Now similar to Eq.\ref{fep}, the $N_{B-L}$ in the prime basis can be written as 
 \bea
N_{B-L}=\sum_{\alpha^\prime}N_{\Delta_{\alpha^\prime}}=-\sum_{\alpha^\prime}\sum_{i}^2\varepsilon_{i\alpha^\prime}\kappa_{i\alpha^\prime}.\label{asyp}
\eea
For numerical purpose, let's assume the total decay parameter $K_i=60, K_{i\mu}=K_{i\tau}=25$. Thus the matrix $V$ which diagonalizes $P_{i\alpha}^0C_{\alpha\beta}$ is given by
\bea
V=\begin{pmatrix}
0.125&0.000&-0.971\\0.701&-0.707&0.166\\0.701&0.707&0.166.
\end{pmatrix}.
\eea
This implies
\bea
\begin{pmatrix}
\varepsilon_{ie^\prime}\\ \varepsilon_{i\mu^\prime}\\ \varepsilon_{i\tau^\prime}\end{pmatrix}=\begin{pmatrix}
0.125&0.000&-0.971\\0.701&-0.707&0.166\\0.701&0.707&0.166.
\end{pmatrix}^{-1} \begin{pmatrix}
\varepsilon_{ie}\\ \varepsilon_{i\mu}\\ \varepsilon_{i\tau}\end{pmatrix}=\begin{pmatrix}
0\\ 1.414\varepsilon_{i\tau}\\ 0\end{pmatrix}.
\eea
Therefore the asymmetry vector in the prime basis is given by
\bea
\begin{pmatrix}
N_{\Delta_{e^\prime}}\\N_{\Delta_{\mu^\prime}}\\N_{\Delta_{\tau^\prime}}\end{pmatrix}=\begin{pmatrix}
0\\ - 1.414\sum_i\varepsilon_{i\tau} \kappa_{i\mu^\prime}\\0
\end{pmatrix}
\eea
which should then be transformed in the unprimed  basis (original basis of leptogenesis) as 
\bea
\begin{pmatrix}
N_{\Delta_{e}}\\N_{\Delta_{\mu}}\\N_{\Delta_{\tau}}\end{pmatrix}=\begin{pmatrix}
0.125&0.000&-0.971\\0.701&-0.707&0.166\\0.701&0.707&0.166.
\end{pmatrix}\begin{pmatrix}
0\\ - 1.414\sum_i\varepsilon_{i\tau} \kappa_{i\mu}\\0
\end{pmatrix}=\begin{pmatrix}
0\\0.996\sum_i\varepsilon_{i\tau} \kappa_{i\mu^\prime}\\-0.996\sum_i\varepsilon_{i\tau} \kappa_{i\mu^\prime}
\end{pmatrix}
\eea
Thus $\sum_{\alpha}N_{\Delta_{\alpha}}=0$ and we have vanishing $N_{B-L}$. However for the proposed $\mu\tau$ mixing scheme, where $P^0_{i\mu}\neq P^0_{i\tau}$ in general, the asymmetry in the muon and tau flavour will couple each other with different coupling strength. Thus even if we consider diagonal $C$ matrix, we obtain a nonvanishing lepton asymmetry. But most interesting point is, when we take the general form of the $C$ matrix (nondiagonal), a portion of the net asymmetry generated due to the interplay of the muon and tau flavour, will be injected in the electron flavour also. 
\begin{figure}[H]
\begin{center}
\includegraphics[scale=.45]{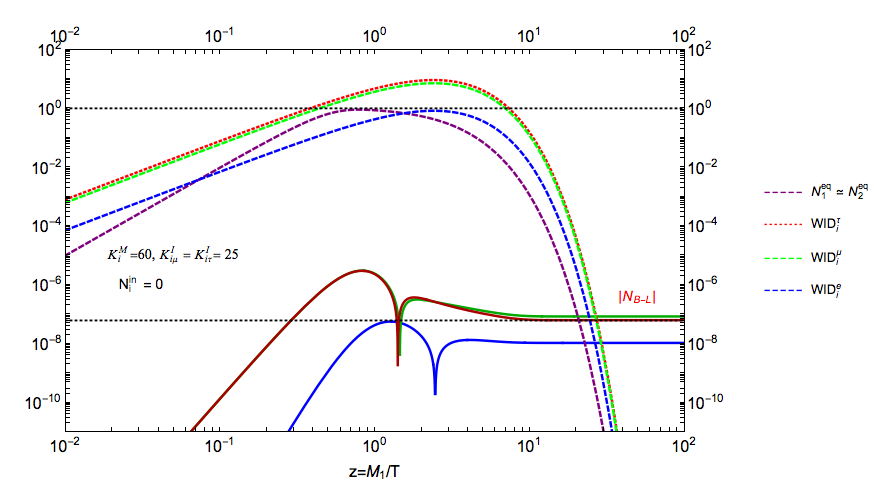} 
\caption{ Variation of $N_{B-L}$ with $z$ assuming $\overline{m}_{\rm max}\frac{4\delta_x}{m^*}=8$ and $|\varepsilon_{i\tau}|\simeq 2.67\times 10^{-5}$. The blue line is the asymmetry injected in the electron flavour through flavour couplings. The green line is net contribution from the muon and tau flavour asymmetries.The red line which matches the observed range after taking into account the injected asymmetry in the electron flavour.}\label{fig7}
\end{center}
\end{figure}
Thus even if we start form a scenario with vanishing production term in the electron flavour ($\varepsilon_{ie}=0$) due to the off-diagonal terms mainly due to the $C_{e\mu}$ and $C_{e\tau}$ term we can generate a nonvanishing lepton asymmetry in the electron flavour. Though, for a fixed value of $\varepsilon_{i\alpha}$, magnitude of the injected asymmetry will depend on how strong the mismatch of the asymmetry in muon and tau flavour.  
 In Fig.\ref{fig7}, along with number densities of the RH neutrinos and flavoured inverse decay rates (dashed lines), we present the variation of $|N_{B-L}|$ (solid lines).  The blue line (solid) represents the injected asymmetry in the electron flavour. The green (solid) represents the net asymmetry generated by muon and tau flavour. The red (solid) line is the final $N_{B-L}$ when we combine all the flavours. This is clear that after taking into account the asymmetry in electron flavour, we obtain correct value of $N_{B-L}$. This shows the importance of the off-diagonal terms in the flavour coupling matrix which are neglected in general in the computation in leptogenesis. The discussed CP extended $\mu\tau$ mixing is thus a novel low energy model   which facilitates the understanding of flavour couplings in Boltzmann equations for leptogenesis in  a very clear way.\\
 
 Before concluding, we would like to highlight the main results of this paper and make few remarks regarding future prospect of this work.\\
 
 $\bullet$ We derive model independent correlations between the Dirac CP phase and the light neutrino mixing angles for  generalized associate $\mu\tau$ symmetries which we name as the associate $\mu\tau$ mixing symmetries. \\
 
 $\bullet$ We have shown that the current data on $\delta$ and $\theta_{23}$ could be better explained by the proposed mixing symmetry.\\
 
  $\bullet$ After a general discussion on $\mu\tau$ mixing which can be realized in many of the neutrino mass models, we discuss the CP extension of it and find novel testable correlations between $\delta$ and the light neutrino mixing angles.\\
  
   $\bullet$  We then discuss the baryogenesis via  leptogenesis mechanism in the three flavour regime and show unlike the CP extended $\mu\tau$ interchange, a resonant leptogenesis is possible in the CP extended $\mu\tau$ mixing and a nonzero baryon to photon ratio always  requires nonmaximal $\theta_{23}$ which is now preferred by the current data.\\
   
 $\bullet$ We have shown  quantitatively, even after inclusion of flavour coupling effect in leptogenesis computation, the usually drawn conclusion of a vanishing asymmetry in the fully flavoured regime is still valid for the CP extended $\mu\tau$ interchange symmetry. \\
 
  $\bullet$  The proposed CP extended $\mu\tau$ mixing is a novel example of a neutrino mass model where the role of  flavour couplings in leptogenesis mechanism is very explicit. \\
   
   The paper is presented entirely from mixing perspective since, whilst the $\mu\tau$ mixing or its CP extensions are very attractive in the light  neutrino mixing sector, they don't entertain  predictions in the mass sector.  Therefore, explicit models with lesser number of parameters with $\mu\tau$ mixing or its CP extended version would be of great interest to search for. Since in that case, besides all the model independent correlations presented in this work, there would be definite statements  on the masses as well. These models would also have interesting consequences in leptogenesis. For example, as we have shown in the CP extended version of $\mu\tau$ mixing, since we don't have the predictions on the masses, we have assumed conditions such as $\cos 2\xi=1$, $K_{i\tau}=25$ etc. But in the models with less  number of parameters, we would have exact statements on the assumed parameters. In that case we can have precise statement on the nonmaximality of $\theta_{23}$ required to generate the observed baryon asymmetry via resonant leptogenesis mechanism even if considering the RH mass scale at $\mathcal{O}$(TeV)\cite{reso,BhupalDev:2012zg}. To constrain the parameter space of such  predictive models, it would be then  interesting to consider the heavy neutrino flavour oscillation effects\cite{Dev:2014laa,Drewes:2016jae,Garbrecht:2018mrp} to increase the robustness of the low energy predictions.

\section{Conclusions}\label{s5}

In this work we promote the idea of $\mu\tau$ mixing symmetry which is a generalized version of the $\mu\tau$ interchange symmetry. After systematically deriving the model independent correlations among the Dirac CP phase and the light neutrino mixing angles in the $\mu\tau$ mixing scheme, we show that the stringent condition of simultaneous maximality of $\delta$ and $\theta_{23}$ which is a well known prediction of the interchange symmetry (currently disfavoured by oscillation data)  could be relaxed. We show that the present data or the current trend on $\delta$ and $\theta_{23}$ seeks deviation from the $\mu\tau$ interchange scheme. Encoding the deviation in a single parameter  $\theta_g$ which generalizes $\mu\tau$ interchange to the  $\mu\tau$ mixing for $\theta_g\neq \pi/4$, we then comment on the strength of the deviation to be consistent with present data and also show how the parameter $\theta_g$ could be related to the model parameters in  neutrino mass models that exhibit $\mu\tau$ mixing at the low energy. Inclusion of the parameter $\theta_g$ opens up  possibilities to explore several classes of the $\mu\tau$ mixing scheme. We particularly discuss the CP extended version of the $\mu\tau$ mixing symmetry and derive novel correlations among $\delta$ and the light neutrino mixing angles. Particularly we show, in  this class of models,  most probable values of $\delta$ prefers maximal Dirac CP violation while the atmospheric mixing angle $\theta_{23}$ is not necessarily maximal. We also show, that unlike  the CP extended version of the $\mu\tau$ interchange, the mixing  scenario is able to explain the observed baryon asymmetry in the three flavour regime via resonant leptogenesis mechanism. Particularly we show, barring a very special choice in the parameter space, the observed baryon asymmetry is proportional to the deviation of $\theta_{23}$ from its maximality. Thus to explain baryon asymmetry simultaneously with neutrino mixing, the CP extended $\mu\tau$ mixing symmetry favours nonmaximal values of $\theta_{23}$. After a qualitative as well  as quantitative comparison of the leptogenesis scenario in the three flavour regime between the CP extended interchange and mixing symmetry, we show,  whilst for the interchange scenario, even if we include  off-diagonal flavour coupling matrix ($C$) in the Boltzmann equation for the leptonic number densities, the usual conclusion of obtaining a vanishing asymmetry in the three flavour regime is unchanged, however, in the mixing scheme, the off-diagonal terms of the $C$ matrix play an important role. In particular, the CP extended $\mu\tau$ mixing  is a novel example of a low energy neutrino mass model, in which even if one does not have any source term for the lepton asymmetry in a particular flavour, a sizeable amount of asymmetry could be injected to that flavour via off-diagonal elements flavour coupling matrix. This in turn emphasizes the importance of the usage of the general structure of  flavour coupling matrix (which is assumed to be diagonal in most of the leptogenesis studies) in the Boltzmann equations.  

\section*{Acknowledgements} 
R. Samanta would like to thank Pasquale Di Bari for a useful discussion related to flavour couplings. R. Samanta would like to thank  the Royal Society, UK and SERB, India for the Newton International Fellowship (NIF). 
\appendix
\section{Appendix}
General expression of $\theta_{23}$ for $\mathcal{G}_1^{g\mu\tau}$ invariance:
\bea
\cos\theta_{23}=\frac{BC\pm \sqrt{B^2C^2-(A^2+B^2)(C^2-A^2)}}{A^2+B^2},
\eea
where we define
\bea
A&=&\sin2\theta_{12}\sin\theta_{13}\cos\delta, \nonumber\\
B&=&\sin^2\theta_{12}-\cos^2\theta_{12}\sin^2\theta_{13}, \nonumber\\
C&=&\cos2\theta_g(\sin^2\theta_{13}+\cos^2\theta_{13}\sin^2\theta_{12}).
\eea
General expression of $\theta_{23}$ for $\mathcal{G}_2^{g\mu\tau}$ invariance:
\bea
\cos\theta_{23}=\frac{BC\pm \sqrt{B^2C^2-(A^2+B^2)(C^2-A^2)}}{A^2+B^2},
\eea
where we define
\bea
A&=&\sin2\theta_{12}\sin\theta_{13}\cos\delta, \nonumber\\
B&=&-\cos^2\theta_{12}+\sin^2\theta_{12}\sin^2\theta_{13}, \nonumber\\
C&=&-\cos2\theta_g(\sin^2\theta_{13}+\cos^2\theta_{13}\cos^2\theta_{12}).
\eea

\end{document}